\documentclass[10pt,a4paper,final]{IEEEtran}
\usepackage{palatino,epsfig,amssymb,cite,xcolor,graphicx}
\usepackage[fleqn]{amsmath}

\newtheorem{prop}{\bfseries Proposition}
\newtheorem{lemma}{\bfseries Lemma}
\newtheorem{theorem}{\bfseries Theorem}

\newtheorem{defn}{\bfseries Definition}
\newtheorem{example}{\bfseries Example}
\def\QED{~\rule[-1pt]{5pt}{5pt}\par}
\newenvironment{proof}{{\em Proof.}}{{ \hfill \QED}}
\newtheorem{remark}{\bfseries Remark}

\newcommand{\hs}{\hspace{4mm}}

\newcommand{\IC}{\mathbb{C}}
\newcommand{\IR}{\mathbb{R}}

\newcommand{\IS}{\mathbb{S}}

\newcommand{\eps}{\varepsilon}

\DeclareMathAlphabet{\matheur}{U}{eur}{m}{n}
\DeclareMathAlphabet{\matheurb}{U}{eur}{b}{n}
\DeclareMathAlphabet{\matheus}{U}{eus}{m}{n}
\DeclareMathAlphabet{\matheuf}{U}{euf}{m}{n}

\newcommand{\RHinf}{\mathcal{RH}_\infty}
\newcommand{\Hinf}{\mathcal{H}_\infty}
\newcommand{\Linf}{\mathcal{L}_\infty}
\newcommand{\RLinf}{\mathcal{RL}_\infty}

\newcommand{\rhol}{{\varrho}}

\newcommand{\AP}{{\cal AP}}
\newcommand{\F}{\mathcal{RF}}
\newcommand{\e}{\mathrm{e}}
\newcommand{\3}{\mathbf{3}_{\epsilon}}

\makeatletter
\def\ps@IEEEtitlepagestyle{
  \def\@oddfoot{\mycopyrightnotice}
  \def\@evenfoot{}
}
\def\mycopyrightnotice{
  \begin{minipage}{\textwidth}
\scriptsize
 \copyright~2023 IEEE. Personal use of this material is permitted. Permission from IEEE must be obtained for all other uses, in any current or future media, including reprinting/republishing this material for advertising or promotional purposes, creating new collective works, for resale or redistribution to servers or lists, or reuse of any copyrighted component of this work in other works.
  \end{minipage}
}

\begin{document}

\title{Exact Instability Margin Analysis and Minimum-Norm Strong Stabilization \\
{\huge -- phase change rate maximization --}
\thanks{This work was supported in part by the Ministry 
of Science and Technology of Taiwan, under grant MOST 110-2221-E-110-047-MY3.}} 
\author{S. Hara,~C.-Y. Kao,~S. Z. Khong,~T. Iwasaki,~Y. Hori 
\thanks{S. Hara is with Global Scientific Information and Computing Center, Tokyo Institute of Technology, 2-12-1 Ohokayama, 
Meguro-ku, Tokyo, Japan. (e-mail: shinji\_hara@ipc.i.u-tokyo.ac.jp)}
\thanks{C.-Y. Kao (corresponding author) is with the Department of Electrical Engineering, 
National Sun Yat-Sen University, Kaohsiung 80421, Taiwan. (e-mail: cykao@mail.ee.nsysu.edu.tw)}
\thanks{S. Z. Khong is an independent researcher. (email: szkhongwork@gmail.com)}
\thanks{T. Iwasaki is with the Department of Mechanical and Aerospace Engineering, University of California at Los Angeles, 
Los Angeles, CA 90095-1597, USA. (e-mail: tiwasaki@ucla.edu)}
\thanks{Y. Hori is with Applied Physics and Physico-Informatics, Keio University,
3-14-1 Hiyoshi, Kohoku-ku, Yokohama, Kanagawa 223-8522, Japan. (e-mail: yhori@appi.keio.ac.jp)}
} %

\maketitle

\begin{abstract}
This paper is concerned with a new %
optimization problem named 
``{\it phase change rate maximization}''
for single-input-single-output %
linear time-invariant %
systems. The problem relates to two control problems, namely robust instability analysis against stable perturbations 
and minimum-norm strong stabilization. We define an index of the instability margin called ``{\it robust instability radius (RIR)}'' 
as the smallest $\Hinf$-norm of a stable perturbation that stabilizes a given unstable system. 
This paper has two main contributions. 
It is first shown that the problem of finding the exact RIR 
via the small-gain condition can be transformed into the problem of 
maximizing the phase change rate at the peak frequency with a phase constraint. 
Then, we show that the maximum is attained by a constant or a first-order all-pass function and 
derive conditions, under which the RIR can be exactly characterized,
in terms of the phase change rate. Two practical applications are provided to illustrate the utility of our results.
\end{abstract}

\begin{IEEEkeywords}
phase change rate maximization, instability analysis, strong stabilization, Nyquist criterion, robust control %
\end{IEEEkeywords}

\section{Introduction}
\label{sec:intro}

Feedback is essential and inevitable for maintaining the desired behaviors against uncertainties in the target systems 
and/or disturbances from the environment as well as for stabilizing an unstable system. Most traditional control 
theories focus on regulation around an equilibrium point or tracking a class of reference signals. Robust control 
theory in particular provides systematic methods for analyzing and synthesizing feedback systems with guaranteed 
stability and control performance in the presence of model uncertainties (see, e.g.,~\cite{zhou:96book}). The most 
fundamental problem is the robust stability analysis: How much uncertainty can be allowed while maintaining 
stability? The answer, called the robust stability radius, has been exactly characterized by the small gain theorem, 
and the robust stability problem has been well understood.

A fundamental counterpart, the robust {\em instability} analysis, is to investigate the maximum allowable perturbation 
for a given unstable feedback system to maintain its instability. The problem has been much less studied so far but is 
of engineering significance. Over the last two decades, feedback control to maintain non-equilibrium state such as 
periodic oscillation has garnered  more recognition as an important design problem for engineering applications including 
robotic locomotion (\cite{grizzle:01,wu:21}) and as an interesting issue in synthetic biology 
(see e.g.,~\cite{Alon2006, Elowitz2000, Kim:IEESB2006}). Robustness of such non-equilibrium states is difficult to 
analyze in general. However, robust instability analysis for linear systems works fairly well for maintaining 
certain classes of nonlinear oscillations as demonstrated in \cite{HIH:LCSS2020, HIH:Automatica2021}, where robust 
oscillations in the sense of Yakubovich are guaranteed by instability of equilibria and ultimate boundedness~\cite{Pogromsky:1999}.
Moreover, as a byproduct of the robust instability analysis, the search for the worst case stable perturbation provides 
a solution to the strong stabilization, a long-standing problem of practical relevance.

Given the above background, the main focus of this paper is the analysis of the instability margin, which poses challenges 
as described below.

{\it (i) Robust Instability Analysis}: 
The problem is similar to but quite different from that of robust stability analysis as pointed out in~\cite{HIH:LCSS2020}, 
which demonstrated through numerical examples why robust instability analysis is far more difficult. Unlike robust stability 
analysis, the small gain condition in terms of the $\Linf$-norm only gives a lower bound of the robust instability 
radius (RIR) (as seen in e.g.,~\cite{Inoue:ECC2013,Tsypkin:Automatica1994}) because existence of a purely imaginary pole 
does not necessarily imply transition from instability to stability (although the opposite transition is implied). Therefore, 
characterization of the RIR requires an intricate analysis to ensure not one but all of unstable poles are perturbed 
to the left half plane. 

{\it (ii) Stabilization by a Minimum-Norm Stable Controller}: 
The difficulty of the RIR analysis can be understood by its equivalence to the minimum-norm strong stabilization 
problem (\cite{HIH:LCSS2020, HIH:Automatica2021}) where a stable controller with minimum norm is sought to stabilize a given 
unstable plant. Although the necessary and sufficient condition for strong stabilizability is known to be the parity 
interlacing property for single-input-single-output (SISO) systems, the required order of a strongly stabilizing controller 
is unknown~\cite{Youla:Automatica1974,Ohta:SICE2001}. It is also unknown whether the problem of strong stabilization 
can be transformed into a convex optimization problem. Hence, no computationally efficient algorithm is known for solving 
such a problem. Minimization of the $\Hinf$-norm on some closed-loop transfer functions has been considered in the literature, 
but only partial solutions have been obtained for the problem in hand due to the difficulty in enforcing the stability
constraint on the controller, e.g.~\cite{Zeren:Automatica2000}. Hence, the requirement of strong stabilization with norm 
constraint on the controller makes the optimization problem extremely difficult.

The instability margin analysis can be formalized by defining the robust instability radius (RIR) in a manner analogous 
to the classical robust stability radius for $\Hinf$-norm bounded perturbations \cite{Hinrichsen:1986}. 
A sufficient condition that allows for exact calculation of the RIR is given in \cite{HIH:Automatica2021}.
The main idea was to find a first-order all-pass function which marginally stabilizes 
a given unstable system, and the paper presented a class of third-order systems to meet the condition 
with an application to the repressilator \cite{Elowitz2000}, showing the effectiveness of this approach.

In this paper, we provide conditions on SISO systems of which
the RIR may or may not be given by the small gain condition, under 
a very mild assumption that the class of systems 
is restricted to those for which the peak gain occurs at a single 
frequency. In the cases where it may, a worst-case
perturbation turns out to be a constant or a first-order all-pass function, 
justifying the aforementioned idea outlined in~\cite{HIH:Automatica2021}.

There are two main theoretical contributions in this paper. 
The first contribution is to show that the problem of finding the exact RIR may be turned into the problem of 
maximizing the phase change rate (PCR) at the peak frequency subject to a phase constraint. 
The fundamental tool to show this is an extended version of the Nyquist criterion for the marginal stability. 
It should be emphasized that the problem of PCR maximization is a completely new  
problem which has not been investigated in the past. 
The second contribution is to provide a complete solution of the maximization problem. 
We will prove that the supremum is attained by a constant or a first-order all-pass function, and 
derive conditions in terms of the PCR, under which the exact RIR is attained. 
It is somewhat surprising that   
no higher-order all-pass function could achieve better PCR than %
a constant or a first-order function. 
This is due to the phase constraint imposed at the peak gain frequency. 
This point will be clarified when we solve the maximization problem. 

To illustrate the applicability of our results, two practical applications are considered. The first one
regards the robustness of the oscillatory behavior of a cyclic gene-regulatory network called ``repressilator.'' 
We apply our results to analyse robustness of the instability of the linearized dynamics against dynamic perturbations. 
The second one regards strong stabilization of a magnetic levitation system, where we demonstrate how our results
can be useful for the design of stable discrete-time controllers with small $\mathcal{H}_\infty$ norms. 

The remainder of this paper is organized as follows. 
Section~\ref{sec:RIR} defines the RIR %
and briefly summarizes results in \cite{HIH:Automatica2021} as a basis for the developments in this paper.
Section~\ref{sec:Characterization} characterizes the
open-loop transfer functions that yield marginally stable closed-loop systems,
based on an extended version of the Nyquist criterion. 
The main results of this paper on the conditions for marginal stabilization and the exact RIR are 
presented in Section~\ref{sec:MainResults}.   
In Section~\ref{sec:SupCR}, we formulate a problem of maximizing the phase change rate and 
provide a solution, which plays an essential role in the proofs of the main results in Section~\ref{sec:MainResults}. 
Two practical applications of our main results are given in Section~\ref{sec:PracticalApp}. 
Proofs of our main results are given in Section~\ref{sec:proof_thm}, just before the concluding 
Section~\ref{sec:Concl} which summarizes the contributions of this paper
and addresses some future research directions.

\noindent 
{\bf Notation:} %
The set of real numbers is denoted by $\IR$. 
$\Re(s)$ and $\Im(s)$ denote the real and imaginary parts of a complex number $s$, respectively. 
The set of proper real rational functions of complex variable $s$ is denoted by $\mathcal{R}_p$.
Let $\Linf$ denote the set of functions that are bounded on the imaginary axis $j\IR$. 
The subset of $\Linf$ which consists of real rational functions that are bounded on $j\IR$ is 
denoted by $\RLinf$. The stable subsets of $\Linf$ and $\RLinf$ are denoted by 
$\Hinf$ and $\RHinf$, respectively. The norms in $\Linf$ and $\Hinf$ are denoted by
$\| \cdot \|_{L_\infty}$ and $\| \cdot \|_{H_\infty}$, respectively.
The open (closed) left and right half complex planes are abbreviated as OLHP (CLHP) and ORHP (CRHP), respectively.

\section{Robust Instability Radius: Definition and Preliminary Results}
\label{sec:RIR}

\medskip

We consider a positive feedback system with perturbed loop transfer function $\tilde h(s)$,
represented by the upper linear fractional transformation (LFT)
\begin{align}
\label{eq:htilde}
 \tilde{h}(s) & = 
 {\cal F}_u \left(
   \begin{bmatrix}
    h_{11}(s) & h_{12}(s) \\ h_{21}(s) &  h(s)
   \end{bmatrix},
   \delta(s)  \right) \\  
 & = h(s)+h_{21}(s)\delta(s)(1-h_{11}(s)\delta(s))^{-1}h_{12}(s), \notag
\end{align}
where $h(s)$ is the scalar nominal loop transfer function and $\delta(s)$ denotes 
the norm-bounded stable perturbation. 
The LFT representation \eqref{eq:htilde} covers a variety of perturbations, including
\begin{eqnarray*}
\mbox{Multiplicative-type} &:& \tilde{h}(s) = (1+ w_m(s)\delta(s)) h(s), 
\\
\mbox{Feedback-type} &:& \tilde{h}(s) = h(s)/(1+ w_f(s)\delta(s)), 
\end{eqnarray*}
where $h_{11}(s)$, $h_{12}(s)$, and $h_{21}(s)$ are respectively set as  
$(h_{11},h_{12},h_{21})=(0,h,w_m)$ and $(-w_f,-w_f,h)$.

We here assume that the nominal feedback system is strictly unstable, i.e., 
the corresponding characteristic equation $1 - h(s) = 0$ has at least one 
root in the ORHP, 
and we consider the problem of determining the minimum norm of $\delta(s)$ 
which makes the feedback system stable. 
Note that the characteristic equation of the perturbed system 
$1 - \tilde{h}(s) = 0$ can be reexpressed as  
\begin{equation} \label{eq:che}
 \delta(s) g(s) = 1 , 
\end{equation}
where $g(s)$ is an unstable transfer function given by 
\[
g(s) := h_{11}(s) + h_{21}(s)h_{12}(s)/\left(1 - h(s)\right). 
\]
Clearly, the weighted sensitivity and complementary sensitivity functions,  
$g_f(s) := w_f(s)/(1 - h(s))$ and $g_m(s) := w_m(s)h(s)/(1 - h(s))$, 
play an important role in the robust instability analysis in the same way as in the robust stability analysis. 
We also note that \eqref{eq:che} presents the %
strong stabilization problem, 
where $g(s)$ and $\delta(s)$ correspond to an unstable plant %
and a stable stabilizing controller to be designed, respectively. 
In what follows, our theoretical investigation %
is based on \eqref{eq:che}.

The rest of the section briefly summarizes some concepts and results from 
\cite{HIH:LCSS2020, HIH:Automatica2021}, which form a basis for the
developments in this paper.
Consider a class of unstable systems defined by 
\begin{equation} \label{eq:G+}
 {\cal G} := \{  g \in \RLinf \; | \;  g \; \mbox{is strictly proper and unstable.} \} . 
\end{equation}
The robust instability radius (RIR) for $g \in {\cal G}$, denoted by $\rho_*(g) \in \IR$, 
with respect to real rational dynamic perturbation $\delta\in \RHinf$, 
is defined as the smallest magnitude of the perturbation that internally stabilizes the system: 
\begin{equation} \label{rho}
\rho_*(g) := \inf_{\delta\in\IS(g)}~\|\delta\|_{H_\infty},
\end{equation}
where $\IS(g)$ is the set of real-rational, proper, stable transfer functions 
internally stabilizing $g$, i.e., 
\begin{equation} \label{IS}
\begin{array}{r}
\hspace{-1mm}
\IS(g) := \{ \delta\in\RHinf:~ \delta(s)g(s)=1 ~ \Rightarrow ~ \Re(s)<0, ~~\\ 
\delta(s)=0, ~ \Re(s)>0 ~ \Rightarrow ~ |g(s)| < \infty ~ \}.
\end{array}
\end{equation}

It is noticed from the well known result on strong stabilizability in \cite{Youla:Automatica1974} 
that $\rho_*(g)$ is finite if and only if the Parity Interlacing Property 
(PIP) %
is satisfied, i.e., the number of unstable real poles of $g$ between any pair of real zeros 
in the closed right half complex plane (including zero at $\infty$) is even. 
Consequently, the class of systems of our interest is defined as
\begin{align}
\begin{split}
\mathcal{G}_n := \{g\in{\cal G} \; | \;  & g \; 
 \mbox{has $n$ unstable poles and} \\
&  \mbox{satisfies the PIP condition.} \}, 
\end{split}\label{eq:classGn}
\end{align}
where $n$ is a natural number. 
We aim to give conditions on $g \in{\cal G}_n$ under which the RIR can be characterized exactly
by the lower bounds given analytically as follows:

\begin{lemma} \rm \cite{HIH:Automatica2021} \;  
\label{prop:lbub}
Let $g \in {\cal G}$ be given. %
Then 
\begin{equation} \label{rhoP}
 \rho_* (g) \geq ~
 \rhol_p:=1/\|g\|_{L_\infty} , \hs \|g\|_{L_\infty} := \sup_{\omega \in \IR} |g(j\omega)| . 
\end{equation} 
Moreover, if $g \in {\cal G}$ has an odd number of unstable poles (counting multiplicity) 
then we have 
\begin{equation} \label{lbo}
 \rho_* (g) \geq \rhol_o := 1/|g(0)|.
\end{equation}
\end{lemma}

\medskip

Let us introduce some notions of stability to facilitate clear and rigorous
presentation of our theoretical developments.
\begin{defn} 
\label{def:MarginalStability} 
\indent
\begin{itemize}
\item
A rational function ${h \in \IR_p}$ is called "exponentially stable" if all the poles of $h$ are in the OLHP.  
\item
A rational function ${h \in \IR_p}$ is called "exponentially unstable" if at least one of the poles of $h$ is in the ORHP.  
\item
A rational function ${h \in \IR_p}$ which is neither exponentially stable nor exponentially unstable is called "marginally stable" 
if any pole of $h$ on the imaginary axis is simple. 
\item A marginally stable rational function $h$ is called "single mode marginally stable" 
if all the poles are located in the OLHP except for either a pole at the origin or 
a pair of complex conjugate  poles on the imaginary axis, say $\pm j\omega_c$. 
To specify the mode on the imaginary axis, the system is called %
$\omega_c$-marginally stable with $\omega_c=0$ for the former and 
$\omega_c\neq0$ for the latter. %
\item
A rational function ${h \in \IR_p}$ which is neither exponentially stable nor exponentially unstable is called  
``polynomially unstable'' if at least one of the poles of $h$ on the imaginary axis is not simple. 
\end{itemize}
\end{defn}

An upper bound on the RIR is obtained as $\|\delta\|_{H_\infty}$ if a stable 
stabilizing perturbation $\delta\in\IS(g)$ is found.
The following Proposition presented in \cite{HIH:Automatica2021} shows that an upper bound 
can always be obtained if marginal stability is achieved with a single mode on the imaginary axis.

\begin{prop} \label{PropS2}  {\em \cite{HIH:Automatica2021} } 
Consider real-rational transfer functions $g$ and $\delta_o$ 
having no unstable pole/zero cancellation between them, 
where the former is strictly proper and the latter is proper and stable (possibly a real constant). 
Suppose the positive feedback system with loop transfer function $\delta_o g$ 
is single mode marginally stable. 
Then, for almost\footnote{
This means that an arbitrarily chosen $\delta_1$ may or may not work to stabilize, 
but when it does not work, a slight modification of it can always make it work.}
any proper stable transfer function $\delta_1$, 
there exists $\eps\in\IR$ of arbitrarily small magnitude $|\eps|$ such that the positive feedback with 
$\delta_\eps:=\delta_o+\eps\delta_1$ internally stabilizes $g$. 
\end{prop}
Note that marginal stability requires that the transfer function $\delta_o$ be chosen to satisfy
\begin{equation} \label{dp}
\delta_o(j\omega_c)=\delta_c := 1/g(j\omega_c), 
\end{equation}
at a critical frequency $\omega_c\geq0$, so that $s=j\omega_c$ is a closed-loop pole. 
If we parametrize a class of perturbations, then $\delta_o$ satisfying
(\ref{dp}) may be determined for each $\omega_c\in\IR$, 
and an upper bound $\|\delta_o\|_{H_\infty}$ on the RIR is obtained when 
the resulting closed-loop poles (i.e., roots of $\delta_o(s)g(s)=1$) are all 
in the OLHP except for $s=\pm j\omega_c$. Note that the exact RIR of $g$ is obtained if
$\|\delta_o\|_{H_\infty}$ coincides with one of the lower bounds such as those stated
in Lemma~\ref{prop:lbub}. In this paper, we will focus on the single mode marginal 
stabilization to derive conditions for getting the exact RIR. 

\section{Open-Loop Characterization of Marginally Stable Closed-Loop Systems}
\label{sec:Characterization}

In this section, we present an extended version of the Nyquist criterion for characterizing single mode 
marginal stability. The result will be used in Section~\ref{sec:MainResults} to derive synthesis conditions on
the certain classes of open-loop systems, which eventually leads to conditions that characterize the exact RIR 
of these systems.

\subsection{Gain/phase change rates}
\label{subsec:PhaseChangeRate}

This section introduces the gain and phase change rates for an $\RLinf$ function,  
which are useful for characterizing the single mode marginal stability. 

For a complex function $f:\IC\rightarrow\IC$, the logarithmic gain and the phase 
angle at point $s\in\IC$ such that $f(s)\neq0$ are defined as $\ln|f(s)|$ 
and $\angle f(s)$, respectively. Here, the angle $\angle f(s)$ of the complex number 
$f(s)$ is not uniquely determined but is chosen so that $\angle f(j\omega)$ is 
continuous on $\omega\in\IR$; such a choice is possible for $f\in{\RLinf}$ 
with no zeros on the imaginary axis. For $f(s)\not=0$, 
$\log f(s):=\ln|f(s)|+j\cdot\angle f(s)$.
Let us denote the logarithmic gain and the phase of frequency
response $f(j\omega)$ by
\begin{equation}
\label{eq:GainPhase1}
A_f(\omega):= \ln|f(j\omega)| , \hs \theta_f(\omega):=\angle f(j\omega).
\end{equation}
In order to characterize how these quantities change when $s=j\omega$ is
perturbed in the direction parallel to the real or imaginary axis,
we consider the logarithmic gain and phase at point $s=\sigma+j\omega\in\IC$, 
and introduce two real functions of two real variables as 
\begin{equation}
\label{eq:GainPhase2}
\hat A_f(\sigma, \omega) := \ln|f(\sigma + j\omega)|, \hs 
\hat \theta_f(\sigma, \omega) := \angle f(\sigma + j\omega).
\end{equation}   
There are four possible directional derivatives $A_f^{\prime} (\omega)$, 
$\theta_f^{\prime} (\omega)$, $M_f^{\prime} (\omega)$, and $\phi_f^{\prime} (\omega)$ 
for a complex function $f \in \RLinf$, which are referred to as 
gain, phase, $\sigma$-gain, and $\sigma$-phase change rates (CRs), respectively, 
and defined as 
\begin{align} 
& A_f^{\prime} (\omega) := \frac{\partial \hat{A}_f}{\partial\omega}(0,\omega),\quad
\theta_f^{\prime} (\omega) := \frac{\partial \hat{\theta}_f}{\partial\omega}(0,\omega),  
\label{eq:GainPhaseCR}\\ 
& M_f^{\prime} (\omega) := \frac{\partial \hat{A}_f}{\partial\sigma}(0,\omega),\quad 
\phi_f^{\prime} (\omega) := \frac{\partial \hat{\theta}_f}{\partial\sigma}(0,\omega) .
\label{eq:S_GainPhaseCR}
\end{align}
Applying the Cauchy-Riemann equations (see, e.g.,~\cite{Ahlfors79}) 
to the complex function $\log f(s)$,  we have the following relations 
among these four change rates.

\begin{lemma} \label{lemma:GPrates} 
For $f \in \RLinf$, the following two relations hold for all $\omega$ satisfying  $f(j\omega) \neq 0$: 
\begin{equation}
\label{eq:CRrelationsGPCRs}
M_f^{\prime} (\omega) = \theta_f^{\prime} (\omega)  , \hs 
\phi_f^{\prime} (\omega)  = - A_f^{\prime} (\omega)  .
\end{equation}
\end{lemma} 

The features and roles of the four CRs are as follows:
\begin{itemize}
\item
The phase CR, $\theta_f^{\prime} (\omega)$, which represents the phase change rate 
along the imaginary axis, plays the most important role in this paper. Since
it has a simple interpretation as the slope of the phase frequency response curve, all
the main theorems in this paper are presented in terms of the phase CR of 
the associated function. 
\item
The $\sigma$-gain CR, $M_f^{\prime} (\omega)$,  which represents the gain change 
rate along the horizontal line parallel to the real axis, is equal to $\theta_f^{\prime} (\omega)$. 
Hence, its importance is the same as that of the standard phase CR.
In general, the $\sigma$-gain CR is easier to handle and compute than the 
phase CR, and hence  the $\sigma$-gain CR is often utilized in the proofs of 
lemmas and theorems in this paper. 
\item
The gain CR, $A_f^{\prime} (\omega)$, which represents the gain change rate 
along the imaginary axis, is instrumental for %
one of the key results. It relates to the phase CR through an %
integral relationships (a counterpart of the well-known Bode's gain/phase integral 
relationships) as shown below.
\item 
The $\sigma$-phase CR, $\phi_f^{\prime} (\omega)$, represents the phase 
change rate along the horizontal line parallel to the real axis. 
It is the negative of the gain CR, and it is the 
least important among the four change rates in the development of main results of this paper.
\end{itemize} 

It should be emphasized that the gain and phase change rates, $A_f^{\prime}(\omega)$ and 
$\theta_f^{\prime}(\omega)$, are not independent for minimum-phase functions. 
We have the following integral relationship linking 
$\theta_f^{\prime}(\omega)$, $A_f^{\prime}(\omega)$, and $A_f(\omega)$.   

\begin{lemma} \label{lem:CR_int_relations} 
{\em (Gain/phase change rate integral relationships)} 
Let $f\in\RHinf$ be a minimum-phase function. For an arbitrary $\omega_p\in\IR$, we have
\begin{align}
\theta_f^{\prime}(\omega_p) = 
\frac{2}{\pi} \int_0^\infty \frac{\omega A_f^{\prime}(\omega)}{\omega^2 - \omega_p^2} \, d\omega.
\label{eq:int_f_1} 
\end{align}
Furthermore, if $\omega_p$ is such that $|f(j\omega_p)|=1$, i.e., $A_f(\omega_p)=0$, we have 
\begin{align}
\theta_f^{\prime}(\omega_p) = 
\frac{2}{\pi}\int_0^\infty A_f(\omega) \frac{\omega^2 + \omega_p^2}{(\omega^2 - \omega_p^2)^2} \, d\omega  . 
\label{eq:int_f_2}
\end{align}
\end{lemma}
\begin{proof}
See Appendix~\ref{appendix:CRint}. %
\end{proof}

The relations \eqref{eq:int_f_1}  and \eqref{eq:int_f_2}  will be used in the proof of Lemma~\ref{thm:bound} 
as one of powerful tools to derive the main theorem of this paper. 

\subsection{Marginal  Stability Criteria via Phase Change Rate}
\label{subsec:CondMS}

We first establish necessary conditions and a necessary and sufficient condition 
for a given positive feedback system with unstable loop transfer function 
being either exponentially stable or marginally stable based on the Nyquist plot 
as a preliminary investigation for our analysis in this paper. 

Let us define $\nu_+ (\cdot)$ (and respectively $\nu_- (\cdot)$) as the number of 
transverse crossing points on the real semi-interval $(1, +\infty)$ from the negative 
imaginary region to the positive one (respectively, from the positive 
to the negative) for the Nyquist plot of $L(j\omega + \cdot)$. Let 
$\nu_o(\cdot):= \nu_+ (\cdot) - \nu_- (\cdot)$.

\begin{lemma} \label{lemma:MS}
Consider a positive feedback system with loop transfer function $L \in {\cal G}_n$. 
The feedback system has all poles in the CLHP %
if and only if there exists $\epsilon_+ > 0$ 
such that the following two equivalent conditions hold  for all $\epsilon \in (0, \epsilon_+)$. 
\begin{itemize}
\item[(i)]
The number of counter-clockwise encirclements of  the Nyquist plot of $L(j\omega + \epsilon)$
about $1+j0$  is equal to $n$, the number of unstable poles of $L$.
\item[(ii)]
$\nu_o(\epsilon)=n$ for the Nyquist plot of $L(j\omega + \epsilon)$. 
\end{itemize}
Moreover, the feedback system is marginally stable if and only if the following two conditions hold
in addition to condition (i) or (ii): (iii-a) $\exists$ $\omega$ such that
$L(j\omega)=1$; (iii-b) If $L(j\omega)=1$, $\left. \frac{d}{ds} L(s) \right|_{s=j\omega} \neq 0$.
\end{lemma}
\begin{proof}
See Appendix~\ref{appendix:MS}. %
\end{proof}

Conditions (i) and (ii), in terms of the ``perturbed'' Nyquist plot
$L(j\omega+\epsilon)$ with small $\epsilon>0$, can be viewed as an extended version of the 
Nyquist criteria for the closed-loop poles being in the CLHP. %
The conditions
are obtained by applying Cauchy's argument principle with a perturbed Nyquist contour 
to deal with closed-loop poles on the imaginary axis. Condition (iii-a) is equivalent to 
the existence of a closed-loop pole on the imaginary axis, while Condition (iii-b) is equivalent to
the imaginary-axis pole(s) being simple. Building on Lemma~\ref{lemma:MS}, the following result %
gives a condition for $\omega_c$-marginal stability.

\begin{prop} \label{prop:OmegaMS2}
Let $\omega_c\geq0$, integer $n\geq1$, and transfer function $L \in{\cal G}_n$ be given.
Consider a positive feedback system with loop transfer function $L$. Suppose
\begin{equation} \label{eq:PeakGain}
A_L'(\omega_c)  =  0 , 
\hs \mbox{i.e., } \hs
\left. \frac{d}{d\omega} |L(j\omega)| \; \right|_{\omega=\omega_c}  = 0 . 
\end{equation}
Then, the feedback system is $\omega_c$-marginally stable if and only if 
condition (i) and one of conditions (ii-a) and (ii-b), indicated below, are satisfied. 
\begin{itemize}
\item[(i)] 
The loop transfer function $L$ satisfies the following:
\begin{equation} \label{eq:ImagPoleFreq}
\begin{array}{rcl}
L(j\omega_c) = 1, &\;& \left. \frac{d}{ds} L(s) \right|_{s=j\omega_c} \neq 0  \\
L(j\omega) \neq 1, &\;& \forall \; \omega \neq \pm \omega_c.
\end{array}
\end{equation} 
\end{itemize}
\begin{itemize}
\item[(ii-a)] 
The Nyquist plot of $L(j\omega)$ satisfies either
$\nu_o(0)=n-1$ if $\omega_c=0$, or $\nu_o(0)=n-2$ if $\omega_c>0$, 
and the phase change rate at $\omega = \omega_c$ is positive, i.e., 
\begin{equation} \label{eq:nonnegativeCR}
\theta_L'(\omega_c) > 0 . 
\end{equation}
\item[(ii-b)] 
The Nyquist plot of $L(j\omega)$ satisfies $\nu_o(0)=n$, 
and the phase change rate at $\omega = \omega_c$ is negative, i.e., 
\begin{equation} \label{eq:nonpositiveCR}
\theta_L'(\omega_c) <  0 . 
\end{equation}
\end{itemize}
\end{prop}
\begin{proof} 
See Appendix~\ref{appendix:OmegaMS2} for a proof. 
\end{proof}

The conditions are obtained by enforcing the extended Nyquist criteria through
the sensitivity analysis at the critical point $L(j\omega_c)=1$, resulting
in the conditions on the phase change rate $\theta_L'(\omega_c)$.

\begin{figure}[h]
\vspace{-2mm}
\centering
\epsfig{file=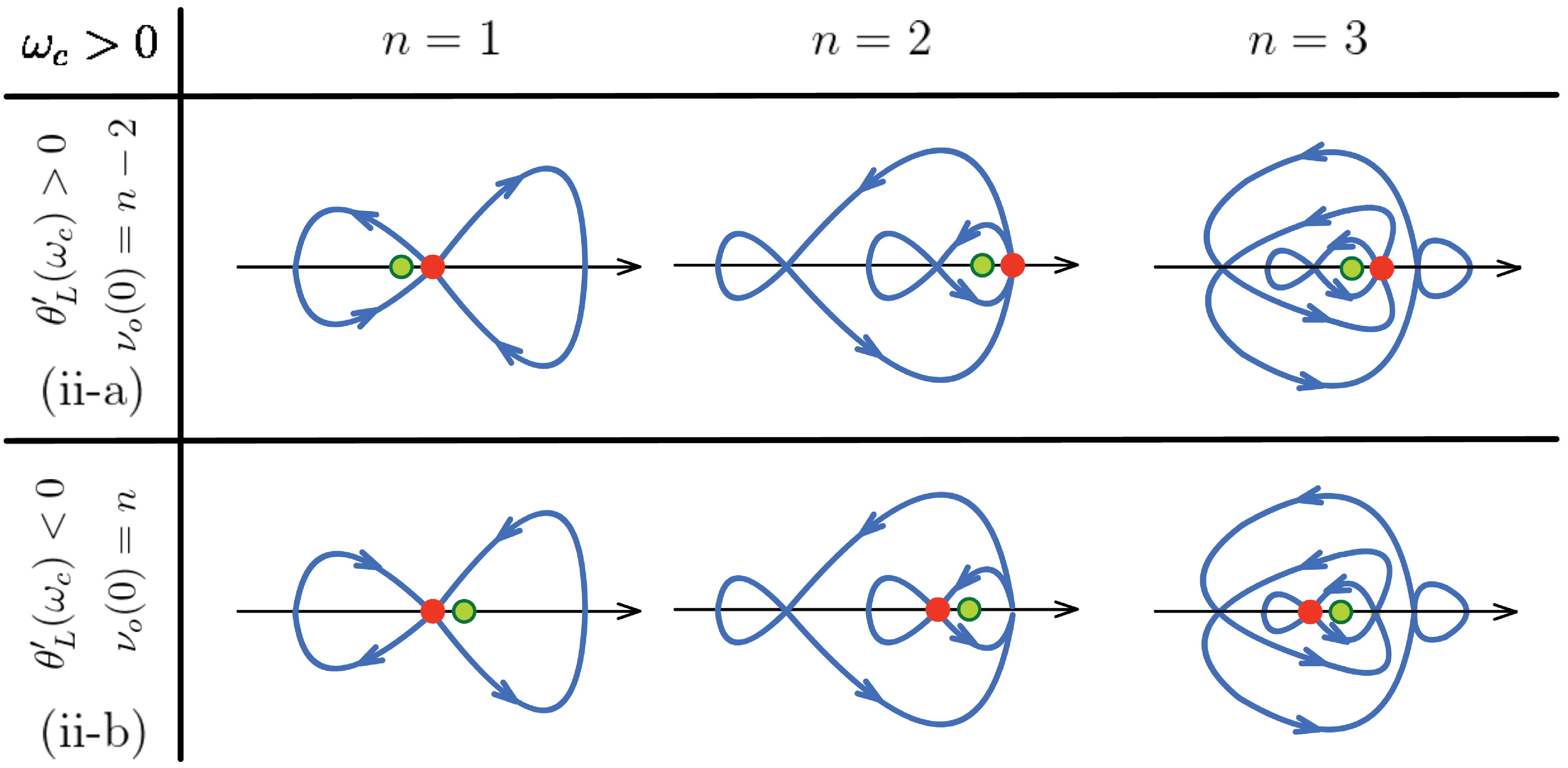,width=80mm} \\
(a) $\omega_c > 0$ \\
\epsfig{file=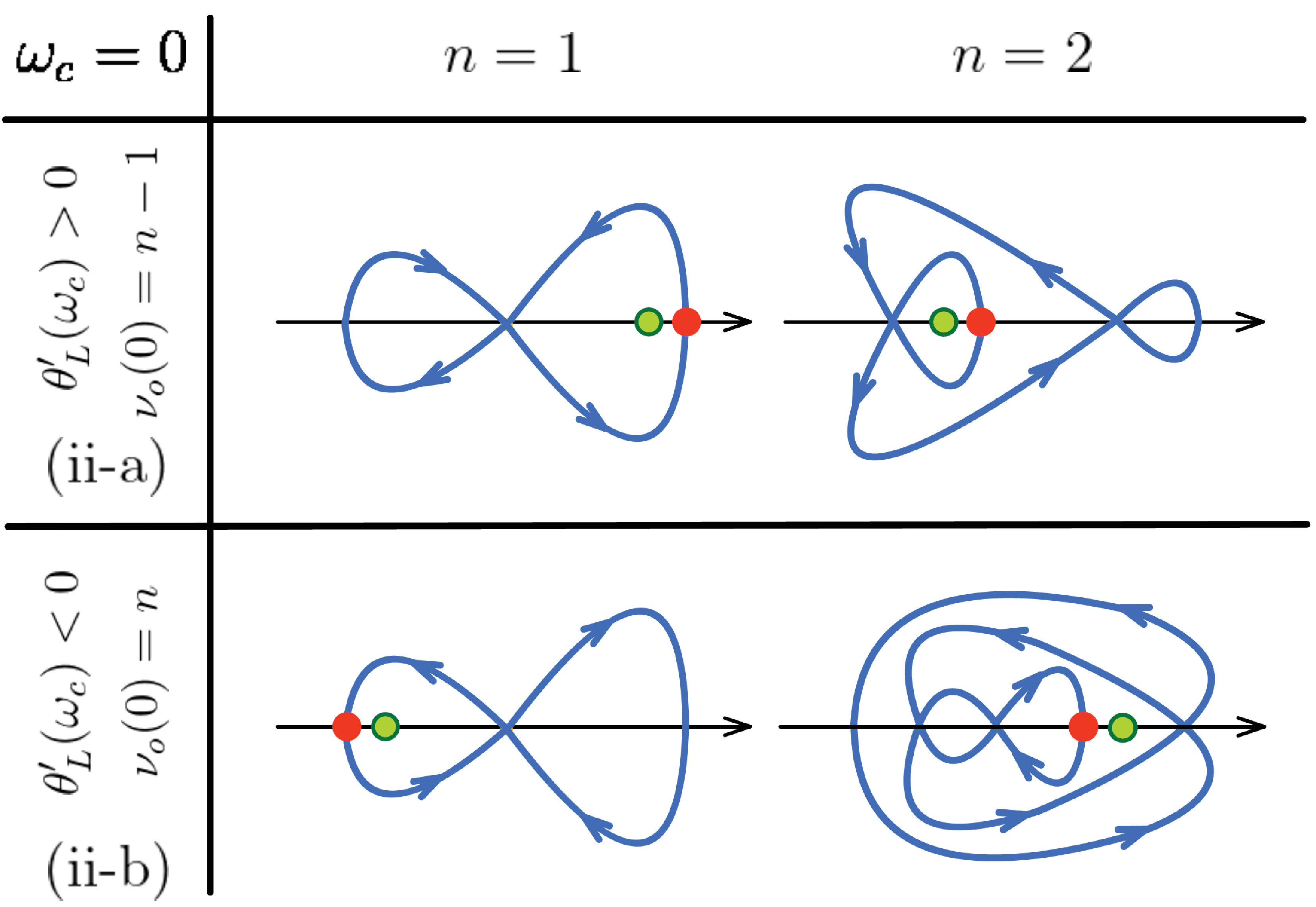,width=60mm} \\
(b) $\omega_c = 0$ \\
\caption{Examples of the Nyquist plot perturbed from $L(j\omega)$ to 
$L(j\omega+\epsilon)$. The blue curve indicates 
the usual Nyquist plot of $L(j\omega)$, where the red dot marks the real axis 
crossing at $L(j\omega_c)=1+j0$. The Nyquist plot of $L(j\omega+\epsilon)$
can also be approximately represented by the blue curve with an appropriate shift 
of the real axis, where the critical point $1+j0$ is now marked by the green dot.
The top figure illustrates the scenario for $\omega_c>0$, while the bottom figure is for $\omega_c=0$.}
\label{fig:ex}
\end{figure}

The idea for the sensitivity analysis can be illustrated by several patterns shown 
in Fig.~\ref{fig:ex}. Under (\ref{eq:ImagPoleFreq}), the Nyquist plot of $L(j\omega)$ 
passes through the critical point $1+j0$ (red dot) at $\omega=\pm\omega_c$ once ($\omega_c=0$)
or twice ($\omega_c>0$). With a small $\epsilon>0$, the Nyquist plot of
$L(j\omega+\epsilon)$ is a slightly perturbed version of $L(j\omega)$ with its shape 
remaining similar, and hence can still be approximately represented by the blue curve 
if the real axis is shifted accordingly. 
When $\theta_L'(\omega_c)$ is positive/negative, the 
real-axis crossing point $L(j\omega_c)$ is perturbed to the right/left to 
become $L(j\tilde{\omega}_c+\epsilon)$ since $\theta_L'(\omega_c)=M_L'(\omega_c)$,
where $\tilde{\omega}_c\approx \omega_c$.
This means that the location of the critical point $1+j0$ relative to the 
perturbed Nyquist plot $L(j\omega+\epsilon)$ with small $\epsilon>0$
can be visualized as the green dot relative to the blue curve in Fig.~\ref{fig:ex}.

Now, for $\omega_c$-marginal stability, $1+j0$ should be encircled $n$ times by 
the perturbed Nyquist plot $L(j\omega+\epsilon)$ (Lemma 4). This happens if
the crossing of $1+j0$ by $L(j\omega)$ at $\omega=\omega_c$ is either (ii-a) 
upward ($\theta'(\omega_c)>0$) with $\nu_o(0)=n-n_c$ where $n_c$ is the 
number of times $L(j\omega)$ passing through $1+j0$, or (ii-b) downward ($\theta'(\omega_c)<0$) 
with $\nu_o(0)=n$ as shown in Fig.~\ref{fig:ex}.

Single model marginal stabilization will be our main focus in the next section,
which is instrumental for obtaining the exact RIR of certain classes of systems.

\section{Main Results}
\label{sec:MainResults}
In this section, Proposition~\ref{prop:OmegaMS2} is applied to find conditions
for marginal stabilizability of $g$ belonging to some subclass of $\mathcal{G}_n$. 
This in turn leads to the exact RIR of $g$.
Specifically, we will consider the subclasses of $\mathcal{G}_n$ where each system has a unique peak frequency.
To this end, we define the following two subsets of ${\cal G}_n$: 
\begin{align}
&\hspace{-0.5cm}\mathcal{G}_n^0:=\{g\in\mathcal{G}_n~|~  
\|g\|_{L_\infty}=|g(0)|>|g(j\omega)|~\forall\omega\neq 0\}, \\
&\hspace{-0.5cm}\mathcal{G}_n^{\#}:=\{g\in\mathcal{G}_n~|~  
\exists~\omega_p>0~\mbox{such that}~\notag \\
&\hspace{1cm}\|g\|_{L_\infty}=|g(j\omega_p)|>|g(j\omega)|~\forall\omega\neq\pm\omega_p
\}.
\end{align}
The assumption on the unique peak frequency is not unreasonable, as most practical systems in reality are expected 
to have such a characteristic. The case where a system has multiple peak frequencies is discussed in Remark~\ref{rmk:multifreq} 
after the presentation of our main results.

\subsection{Conditions for marginal stabilization}
\label{subsec:MainResultMS}

The following theorem provides necessary and sufficient conditions for marginal stabilization 
of system $g$ in ${\cal G}_n^0$ or ${\cal G}_n^{\#}$, by some system $f$ satisfying 
$\|f\|_{H_\infty}=1/\|g\|_{L_\infty}$. As $1/\|g\|_{L_\infty}$ is a lower bound on the RIR of $g$, single mode 
marginal stabilization of a $g$ by such $f$ implies that $g$ has the exact RIR equal to $1/\|g\|_{L_\infty}$ by 
Proposition~\ref{prop:lbub}.  

\begin{theorem} \label{thm:Main} 
\hs
\begin{itemize}
\item[(I)]  
Given $g \in {\cal G}_n^0$,  
$g$ can be marginally stabilized by a stable system $f$ 
with $\|f\|_{H_\infty} = 1/\|g\|_{L_\infty} = 1/|g(0)|$ if and only if $n=1$ and 
\begin{align} \label{eq:CR0}
\theta_g'(0) > 0.
\end{align}
\item[(II)] Given $g \in {\cal G}_n^\#$ for which 
the peak gain occurs at $\omega_p$,
 $g$ can be marginally stabilized by a stable system $f$ 
with $\|f\|_{H_\infty} = 1/\|g\|_{L_\infty} = 1/|g(j\omega_p)|$ if and only if $n=2$ and
\begin{align} \label{eq:CRomega}
 \theta_g'(\omega_p)    >  \left|\sin(\theta_g(\omega_p))/\omega_p\right|.
\end{align}
\end{itemize}
\end{theorem}
\begin{proof}
For the sake of readability, we postpone the full proof to~section~\ref{subsec:ProofMainTheorem}.
\end{proof}

Let us briefly explain the main idea behind the two stabilizability conditions (\ref{eq:CR0}) and (\ref{eq:CRomega}). 
These conditions are obtained by applying Proposition~\ref{prop:OmegaMS2}
to the positive feedback system with loop transfer function $L:=gf$. 
Due to the norm constraint on $f$, the open-loop transfer function $L$ must have peak gain $\|L\|_{L_\infty}=1$
at frequency $\omega_p$, which leads to condition (ii-a) of Proposition~\ref{prop:OmegaMS2}, and 
condition~\eqref{eq:nonnegativeCR} is equivalent to $\theta_g'(\omega_p)+\theta_f'(\omega_p)>0$. 
We can also see that condition (ii-b) does not hold. 
Hence, $g$ is marginally stabilizable by some $f$ when $\theta_g'(\omega_p)$ is larger than the infimum 
of $-\theta_f'(\omega_p)$ over the set of all suitable $f$, given in (\ref{eq:CR0}) and (\ref{eq:CRomega}).

More accurately, stabilizability conditions (\ref{eq:CR0}) and (\ref{eq:CRomega}) emerge from the following phase
change rate maximization problem:
\begin{align} \label{eq:GPCR}
\sup \ & \theta_f'(\omega_p) \mbox{ subject to } f\in\RHinf, \ f(j\omega_p)g(j\omega_p)=1,  \notag \\ 
&
 |f(j\omega)| \leq |f(j\omega_p)| = 1/\|g\|_{L_\infty} \; ; \; \forall \omega \in \IR . 
\end{align}
We make the following claim.

\vspace{0.1cm}
\noindent
{\bf Claim:} The maximum of the phase change rate maximization problem~\eqref{eq:GPCR} is equal to 0 when $\omega_p=0$, 
and $-|\sin(\theta_g(\omega_p))|/|\omega_p|$ when $\omega_p\not = 0$. 

\vspace{0.1cm}
Section~\ref{sec:SupCR} is devoted to solving the phase change rate maximization problem~\eqref{eq:GPCR} and proving the above claim;
see Theorem~\ref{thm:supHinf}. Meanwhile, several remarks on Theorem~\ref{thm:Main} are in order. 

\begin{remark}
Note that the norm constraint on $f$ implies that the loop transfer function $L$ has a single peak gain at $\omega_p$.
Therefore, the marginal stability stated in Theorem~\ref{thm:Main} is single mode. That is, 
there is only one ``marginally stable'' pole (at the origin) for case (I), and a pair of ``marginally stable'' poles
(at $\pm j\omega_p$) for case (II). 
\end{remark} 

\begin{remark}
Note that $\sin(\theta_g(\omega))/\omega\to\theta_g'(0)$ as $\omega\to 0$. Therefore, inequality~\eqref{eq:CR0} is not 
simply the limiting case of~inequality~\eqref{eq:CRomega}. 
\end{remark}

\begin{remark}
\label{rmk:polyuns}
For $g\in\mathcal{G}_n^0$ or $\mathcal{G}_n^{\#}$ with $n\ge 3$, it may be possible to find a stable $f$ 
with $\|f\|_{H_\infty} = 1/\|g\|_{L_\infty}$ such that all poles
of the feedback system are in the CLHP. However, since the loop transfer function $L$ has a single peak gain
as a result of the norm constraint on $f$, the poles on the imaginary axis must be located at $0$ or $\pm j\omega_p$,
and hence must have multiplicity larger than one. In such cases, the closed-loop system is polynomially unstable. 
\end{remark}

\vspace{-0.2cm}

\subsection{Conditions for exact RIR}
\label{subsec:MainResultExactRIR}

Based on Theorem~\ref{thm:Main} and Proposition~\ref{prop:lbub}, conditions
for attaining the RIR of a system exactly given by the small gain condition are derived. 
Moreover, we also show that for some systems, their RIR's are not given
by the small gain condition.

\begin{theorem}\label{thm:exactRIR}
\hs
\begin{itemize}
\item[(I)]
(Necessity)
Let $g\in{\cal G}_n^{s}:={\cal G}_n^0\cup{\cal G}_n^{\#}$ for some positive integer $n$, and $\omega_p$ be its peak-gain frequency. 
Define
\begin{align}
\label{eq:mu_g}
\mu_g(\omega_p):=\begin{cases}
0 & \mbox{if $\omega_p=0$;}\\
\left|\sin\left(\theta_g(\omega_p)\right)/\omega_p\right| & \mbox{otherwise.}
\end{cases}
\end{align}
If $\rho_*(g) = 1/\|g\|_{L_\infty}$, then $\theta_g'(\omega_p) \ge \mu_g(\omega_p)$.
\item[(II)] 
(Sufficiency)
Let $g\in{\cal G}_1^0\cup{\cal G}_2^{\#}$,  and $\omega_p$ be its peak-gain frequency. 
Then $\rho_*(g) = 1/\|g\|_{L_\infty}$ if
\begin{align}
\label{cond:exactRIRa}
\theta_g'(\omega_p) > \mu_g(\omega_p),
\end{align} 
where $\mu_g(\cdot)$ is defined in~\eqref{eq:mu_g}.
\item[(III)] Let $g\in{\cal G}_n^{\#}$, where $n$ is a positive odd integer. For such $g$, we have 
$\rho_*(g) > 1/\|g\|_{L_\infty}$.
\end{itemize}
\end{theorem}
\begin{proof} See Section~\ref{subsec:ProofExactRIR} for a proof. \end{proof} 

Several remarks are in order. 

\begin{remark}
For $g\in{\cal G}_1^0\cup{\cal G}_2^{\#}$, the gap between the necessary and the sufficient conditions is due to the  
cases where the exact RIR is attained through the polynomial instability. For example, consider the reduced-order model 
for the magnetic levitation system $g_r(s)=k/(-s^2+p^2)$ (see Section~\ref{sec:meglev_sys}). One can readily verify
that $g_r\in{\cal G}_1^0$, $\|g_r\|_{L_\infty}=|g_r(0)|=k/p^2$, and $\theta_{g_r}'(0)=0$. It is shown in~\cite[Section 3]{IFAC23_KHetal}
that the exact RIR of $g_r$ is $1/\|g_r\|_{L_\infty}$, and the critical controller results in a closed-loop system with double 
poles at the origin. This example shows that~\eqref{cond:exactRIRa} is not necessary. 
\end{remark}

\begin{remark}
For $g\in\mathcal{G}_n^{0}$ with $n\ge 2$, or $g\in\mathcal{G}_{2m}^{\#}$ with $m\ge 2$, 
the RIR of $g$ may be given by the
small gain condition. In these cases, one must find a critical controller with $\Hinf$ norm equal to  
$1/\|g\|_{L_\infty}$ such that the closed-loop system has all its poles in the CLHP, and multiple poles 
at the origin (for $g\in\mathcal{G}_{n}^0$), or $\pm j\omega_p$ (for $g\in\mathcal{G}_{2m}^{\#}$). 
The reasoning for this fact is similar to that given in Remark~\ref{rmk:polyuns}.
\end{remark}

\begin{remark}
\label{rmk:multifreq}
For $g\in\mathcal{G}_1$ or $\mathcal{G}_2$ with multiple peak frequencies, it can be single mode marginally stabilized by 
some $f$ with $\Hinf$-norm arbitrarily close to $1/\|g\|_{L_\infty}$, and hence has the exact RIR, if 
\begin{itemize}
\item for $g\in\mathcal{G}_1$, $|g(0)|=\|g\|_{L_\infty}$ and inequality~\eqref{eq:CR0} holds; 
\item for $g\in\mathcal{G}_2$, there exists one peak frequency where inequality~\eqref{eq:CRomega} holds.
\end{itemize}
For such $g$, one can apply an inverse notch filter which maintains 
the gain at the frequency where~\eqref{eq:CR0} or~\eqref{eq:CRomega} holds and decreases the gain at all other frequencies 
by an appropriately small amount. This way, the filtered $g$ has a unique peak frequency for which Theorems~\ref{thm:Main}
and~\ref{thm:exactRIR} become applicable. This in turn leads to the aforementioned results. 
For details, please refer to~\cite[Section~6]{arX22_HKetal}.
\end{remark}

\begin{example}
We here apply Theorem~\ref{thm:exactRIR} to a class of second-order unstable systems 
$g \in {\cal G}$ represented by 
\begin{equation} \label{g2nd}
g(s)=1/(s^2+ps+q) \; ; \; q \neq 0
\end{equation} 
to show the effectiveness of the results.  

First note that $g \in {\cal G}_1$ if and only if $q<0$; 
$g \in {\cal G}_2$ if and only if $q>0$ and $p<0$.
Also note that $|g(j\omega + \epsilon)|^2 := 1/D_\epsilon(\Omega)$, 
where $\Omega:=\omega^2$,
\begin{align*}
D_\epsilon(\Omega):=
(\Omega^2 + (p^2-2q)\Omega + q^2) + 2p(\Omega + q)\epsilon + o(\epsilon),
\end{align*}
and $o(\epsilon)$ represents the terms with $\epsilon$ of second or higher orders.
The peak gain frequency $\omega_p$ of $g$ can be found by minimizing
$D_0(\Omega)$ subject to the constraint $\Omega\in[0,\infty)$. One can readily verify
that the minimum of $D_0(\Omega)$ occurs at $\Omega_p:=q-p^2/2$ when $q\ge p^2/2$, 
and at $\Omega_p=0$ otherwise. Thus, we conclude that
\begin{itemize}
\item for any $g\in\mathcal{G}_1$, $\omega_p = 0$ because $q-p^2/2$ must be negative; 
i.e., $g \in {\cal G}_1$ implies $g \in {\cal G}_1^0$.
\item for $g\in\mathcal{G}_2$, $\omega_p = 0$ when $2q \le p^2$; otherwise, $\omega_p>0$
and $\omega_p^2 = q - p^2/2 = \Omega_p$. 
\end{itemize}
Now the $\sigma$-gain change rate can be calculated based on the form of \eqref{eq:S_GainPhaseCR}. 
One can verify that
\begin{equation} \label{eq:sigmaGCR}
M_g'(\Omega) = - p(\Omega + q)/\left(\Omega^2 + (p^2-2q)\Omega + q^2\right), 
\end{equation} 
and we have $M_g'(0) = - p/q$ and $M_g'(\Omega_{p}) = - 2/p$. 
Applying Theorem~\ref{thm:exactRIR}, we have the following results. 
\begin{itemize}
\item[(I)] 
Suppose $g \in {\cal G}_1$, i.e., $q<0$. 
Then, $\rho_*=\varrho_o:=1/|g(0)| = |q|$ holds if and only if  $p \geq 0$.
\item[(II)] 
For any $g \in {\cal G}_2^\#$, i.e., $p<0, \; 2q > p^2 > 0$, we have 
$\rho_*=\varrho_p:=1/\|g\|_{L_\infty} = 1/|g(j\omega_{p})|$ 
where $\omega_p^2 = \Omega_{p} = q - p^2/2$. 
\end{itemize}

The proofs of the results (I) and (II) are as follows. 
Regarding (I), $p \geq 0$ is equivalent to $M_g'(0) = - p/q \geq 0$. 
Although statement (I) of Theorem~\ref{thm:exactRIR} only gives the necessity for $p=0$, we can readily show that 
a constant $\delta$ marginally stabilizes even for $p=0$, which gives rise to the sufficiency. 
This completes the proof of (I). 
We compare the values of $|\sin(\theta_g(\omega_p))/\omega_p|$ 
with $-2/p$ to show (II). 
We can show that $|\sin(\theta_g(\omega_p))/\omega_p|^2 = 4/(4q-p^2) < 4/p^2  = M'_g(\Omega_p)^2$, 
since $2q > p^2$ is equivalent to $4q-p^2 > p^2$.  
This completes the proof of (II). 

For $g \in {\cal G}_2^0$, i.e., the case where 
$p<0, \;  0 < 2q \leq p^2$, we do not yet know whether $g$ has the exact RIR.  
\end{example}

\section{Supremum of Phase Change Rate over Stable Transfer Functions}
\label{sec:SupCR}

In this section, the phase change rate maximization problem~\eqref{eq:GPCR} is solved and 
the claim stated in the previous section is proven. {  It will be shown that the supremums
are attained by the zeroth-order and the first-order all-pass functions for $\omega_p=0$ and
$\omega_p\not = 0$, respectively.} 

\subsection{Phase change rate maximization}
\label{subsec:CRmaximProblem}
To solve the optimization problem described in~\eqref{eq:GPCR}
for $\omega_p\ge 0$, notice that the constraint $g(j\omega_p) f(j\omega_p) = 1$ is equivalent to the gain 
and phase conditions:
\begin{equation}
\label{eq:GainPhaseCond}
 A_f(\omega_p) \cdot A_g(\omega_p) = 0, \; \; 
\theta_f(\omega_p) + \theta_g(\omega_p)  = 0. 
\end{equation}
Also notice that the phase variation $\theta_f'(\omega)$ of a function $f$ is invariant to a 
constant scaling on $f$. Therefore, the essential aspects of the constraints in~\eqref{eq:GPCR} are
the phase of $f$ at $\omega_p$, and that $f$ attains its $\Hinf$-norm at $\omega_p$. The 
magnitude of $f$ at $\omega_p$ is immaterial because the first equation in~\eqref{eq:GainPhaseCond} can always 
be satisfied via a constant scaling on $f$. As such, without loss of generality, we will assume
that the functions under consideration in this section have unit norm. To facilitate development, let us
introduce two classes of transfer functions $f(s)$ based on its frequency response  
$f(j\omega) = |f(j\omega)| e^{j\theta_f(\omega)}$. Define 
\begin{align*}
 \F_{\omega_p, \theta_p}  := \{ f  \in \RHinf \; | \; 1=\|f\|_{H_\infty}=&|f(\omega_p)|, \\ 
& \theta_f(\omega_p) = \theta_p \}. 
\end{align*}
This represents a class of proper real rational stable functions of which 
the $\Hinf$-norm is attained at a specified frequency $\omega_p$ and the phase at that frequency is 
equal to a specified value $\theta_p$. Moreover, let us also consider 
\begin{align*}
\AP_{\omega_p,\theta_p}:= \AP \cap\F_{\omega_p,\theta_p},
\end{align*}
where $\AP := \{f\in\RHinf \; | \;  |f(j\omega )|  = 1, \forall \omega \}$
is the set of all stable real rational all-pass functions with unit norm.
Note that $\AP$ includes $0^{\rm th}$-order all-pass functions, 
which take values $\pm 1$ over the entire complex plane.  
Clearly, $\AP_{\omega_p,\theta_p}$ is a subset of $\AP$ containing all-pass functions whose phase at $\omega_p$ is 
constrained to be a specified value $\theta_p$. 

Using $\F_{\bullet,\bullet}$, we consider the following equivalent maximization problems of~\eqref{eq:GPCR}:

\smallskip
\noindent 
{\bf [phase change rate maximization]}
\begin{align}
\label{eq:sameSol}
\sup_{f\in \F_{\omega_p, -\theta_g(\omega_p)}} \theta_f'(\omega_p) \equiv 
\sup_{f\in \F_{\omega_p, -\theta_g(\omega_p)}} M_f'(\omega_p).
\end{align}
The equivalence of the two problems follows straightforwardly from Lemma~\ref{lemma:GPrates}.
Also note that not only the two problems have the same supremum, but the arguments of supremum are also identical. 

To solve~\eqref{eq:sameSol} (with $-\theta_g(\omega_p)$ replaced by $\theta_p$ for notational simplicity), 
{  we will show that the supremum of $\theta_f'(\omega_p)$ and $M_f'(\omega_p)$ over $\F_{\omega_p,\theta_p}$ 
is in fact the same as that over $\AP_{\omega_p,\theta_p}$. This follows from a key 
observation that the minimum-phase factor of a stable function does not ``help'' in elevating the phase change 
rate at the peak frequency. To this end, let us consider the following problem whose solution is a lower bound for~\eqref{eq:sameSol}
\begin{align}
\label{eq:Problem_AP}
\sup_{f\in \AP_{\omega_p, \theta_p}} \theta_f'(\omega_p) \quad \equiv \quad
\sup_{f\in \AP_{\omega_p, \theta_p}} M_f'(\omega_p). 
\end{align}
}
\begin{prop}  \label{thm:AP} 
Consider the optimization problem~\eqref{eq:Problem_AP}. 
\begin{itemize}
\item[(I)] For $\omega_p=0$, we must have $\theta_p\in\{0,\pi\}$ (mod $2\pi$). In this case, 
$$
\sup_{f\in \AP_{0,\theta_{p}}} \theta_f'(0) =  \sup_{f\in \AP_{0,\theta_{p}}} M_f'(0) = 0.
$$
The supremum is attained by $f(s)= 1$ or $f(s)= -1$.
\item[(II)] For $\omega_p \not= 0$ and $\theta_{p}\in(-\pi,\pi]$ (mod $2\pi$), we have 
\begin{equation} \label{eq:supAP}
\hspace{-3mm}
\sup_{f\in \AP_{\omega_p,\theta_{p}}} \theta_f'(\omega_p) =  \sup_{f\in \AP_{\omega_p,\theta_{p}}} M_f'(\omega_p) = 
-\left|\frac{\sin(\theta_{p})}{\omega_p}\right|.
\end{equation}
Moreover, when $\theta_{p}\not\in\{0,\pi\}$, the supremum is attained by the first-order all-pass function 
of the form $f(s)=\frac{a-s}{a+s}$ or $f(s)=\frac{s-a}{a+s}$ satisfying 
the phase constraint $\theta_f(\omega_p) = \theta_{p}$. 
When $\theta_{p}\in\{0,\pi\}$, the supremum is attained by a zeroth-order all-pass functions; 
i.e., $f(s)= 1$ or $f(s)= -1$. 
\end{itemize}
\end{prop}

{ 
Furthermore, the key observation that leads to the solution of~\eqref{eq:sameSol} can be formulated as follows.
\begin{prop} \label{prop:FR=AP}
Optimization problems~\eqref{eq:sameSol} and~\eqref{eq:Problem_AP} have the same solutions. That is,
\begin{align}
\label{eq:FR=AP}
\sup_{f\in \F_{\omega_p, \theta_p}} \theta_f'(\omega_p) 
= \sup_{f\in \AP_{\omega_p, \theta_p}} \theta_f'(\omega_p).
\end{align}
\end{prop}

Proofs of Propositions~\ref{thm:AP} and~\ref{prop:FR=AP} can be found in Sections~\ref{subsec:ProofAP} 
and~\ref{subsec:ProofSupHinf}, respectively. With these two propositions, we arrive at the main
result of this section. 
}
\begin{theorem}  \label{thm:supHinf} 
Consider the optimization problem~\eqref{eq:sameSol}. 
\begin{itemize}
\item[(I)]  
For $\omega_p=0$, we must have $\theta_p\in\{0,\pi\}$ (mod $2\pi$). In this case, 
\begin{equation} \label{eq:supHinf1}
  \sup_{f \in \F_{\omega_p, \theta_p}} \theta_f^\prime (\omega_p) = 0.
\end{equation}
\item[(II)]  
For $\omega_p \not= 0$ and $\theta_{p}\in(-\pi,\pi]$ (mod $2\pi$), we have 
\begin{equation} \label{eq:supHinf2}
  \sup_{f \in \F_{\omega_p, \theta_p}} \theta_f^\prime (\omega_p)  
  =  -\left|\sin(\theta_p)/\omega_p\right|.
\end{equation}
\end{itemize}
Moreover, the supremum is attained by $f(s)=1$ or $f(s)=-1$ when it is zero.  
When the supremum is not zero, it is attained by a first-order all-pass function
as described in statement (II) of Proposition~\ref{thm:AP}. 
\end{theorem} 
\begin{proof}
The result follows from Propositions~\ref{thm:AP} and~\ref{prop:FR=AP} in a straightforward manner. 
\end{proof}

One may find the results in Proposition~\ref{thm:AP} and Theorem~\ref{thm:supHinf} 
somewhat counter-intuitive, as one may think a higher-order all-pass function would give
better result due to more optimizable parameters it provides. This additional ``freedom'' 
is not useful because of the constraint on the phase at $\omega_p$. To better understand 
this point, let us compare the phase change rate of a generic second-order all-pass function 
with that of the first-order. 

\begin{example}
We here evaluate the phase change rate of the second-order all-pass function 
\begin{align*}
f(s) := \frac{a-bs+s^2}{a+bs+s^2}, \; a, \ b>0.  
\end{align*}
Since the phase and the $\sigma$-gain change rates are the same, we calculate the latter instead.
For given $\theta_p$ and $\omega_p$, the phase constraint $\theta_f(\omega_p)=\theta_p$ requires that  
the parameters $a$ and $b$ be chosen so that 
\begin{align}
\sin (\theta_p) = -\frac{2(a - \omega_p^2)b\omega_p}{(a - \omega_p^2)^2 + b^2\omega_p^2} 
\label{eq:sinTp}
\end{align} 
holds. Furthermore, one can verify that
\begin{align}
\hspace*{-5mm} 
M_{f}'(\omega)
= - \frac{2(a - \omega^2)b}{(a - \omega^2)^2 + b^2\omega^2} 
  - \frac{4\omega^2b}{(a - \omega^2)^2 + b^2\omega^2}.
\label{ex2:ineq1}
\end{align} 
For $\omega_p\not = 0$, substituting~\eqref{eq:sinTp} into equation~\eqref{ex2:ineq1} yields
\begin{align}
 {  M_{f}'(\omega_p)}  
= \frac{\sin(\theta_p)}{\omega_p}+\frac{2\omega_p^2}{a-\omega_p^2}\frac{\sin (\theta_p)}{\omega_p}.
\label{ex2:ineq2}
\end{align}
Suppose $\sin(\theta_p) = 0$. This implies $a-\omega_p^2=0$, and~\eqref{ex2:ineq1} implies
$M_{f}'(\omega_p) <0$. If $\sin(\theta_p)/\omega_p < 0$, then $a-\omega_p^2>0$ and clearly~\eqref{ex2:ineq2} 
implies $M_{f_2}'(\omega_p)<-|\sin(\theta_p)/\omega_p|$. Finally, 
$M_{f_2}'(\omega_p)<-|\sin(\theta_p)/\omega_p|$ also holds when $\sin(\theta_p)/\omega_p > 0$. To see 
this, note that in this case $a-\omega_p^2<0$ and one can verify that $1+2\omega_p^2/(a-\omega_p^2) < -1$. 
Thus we conclude that the best first-order all-pass function is better than 
any second-order one, as the phase change rate of the first-order all-pass function with the same
constraint is equal to $-|\sin(\theta_p)/\omega_p|$.
\end{example}

\section{Practical Applications}
\label{sec:PracticalApp}

In this section, we apply our main results to analyze (in)stability properties of system models that are derived from
real-world applications. Section~\ref{sec:repressilator} is concerned with exact robust instability analysis for a 
biological network oscillator called ``repressilator'', of which the linearized model used is in ${\cal G}_2^{\#}$.  
Section~\ref{sec:meglev_sys} is about digital stable controller synthesis for a magnetic levitation system, where 
the model belongs to ${\cal G}_1^0$. The goal is to illustrate that our theoretical results are applicable to problems 
of significance to provide useful information. Note that the problem settings here are more practical than those 
considered in \cite{IFAC23_KHetal}. 

\subsection{Robust Instability Analysis for Repressilator}
\label{sec:repressilator}

Consider the repressilator with three dynamical units in a cyclic loop~\cite{Elowitz2000}.  
Its linearized model around an equilibrium state is approximated by a positive feedback system 
with loop-transfer function represented by 
\begin{align*}
h_e(s) = \frac{-k_e D^{\tau}(s)}{(s+\alpha_1)(s+\alpha_2)(s+\alpha_3)},
\end{align*}
where $k_e>0$ depends on the equilibrium state of the original nonlinear system~\cite{HIH:Automatica2021} 
and $D^{\tau}(s)$ denotes a  Pad\'{e} approximation of the time-delay transfer function $\mathrm{e}^{-\tau s}$.
The nominal system with the characteristic equation $1=h_e(s)$ is assumed to be exponentially unstable.
The subscript $\bullet_e$ is used to indicate that the quantity $\bullet$ depends on the equilibrium state, which is subject to 
the DC-gain of the uncertainty $\delta(s)$ denoted by $e  := \delta(0)$.
For more details about the repressilator model, see~\cite{HIH:Automatica2021}. 

We are interested in assessing robust instability against a ball type multiplicative perturbation 
with a frequency weight function $w_m(s) := (1 + \zeta T_m s)/(1 + T_m s)$. 
As explained at the beginning of Section~\ref{sec:RIR},   
the corresponding characteristic equation is $1-\delta(s) g_e(s) = 0$, where $g_e(s) = w_m(s) h_e(s)/(1-h_e(s))$. 
We set the parameters of the nominal system be the same as in \cite{HIH:Automatica2021}, i.e.,  
$\alpha_1=0.4621$, $\alpha_2=0.5545$, and $\alpha_3=0.3697$, 
and the fifth-order Pad\'{e} approximation is used for the delay with $\tau = 0.25$, 
which mainly accounts for maturation time of protein. 
The weight function $w_m(s)$ is defined with $T_m = 0.3$ and $\zeta = 10$.

Since $g_e$ is parametrized by the DC-gain of its stabilizing perturbation $\delta$, obtaining
its instability margin requires a more elaborated analysis. For a given $e$, we apply Lemma~4 
of~\cite{HIH:Automatica2021} to verify whether $g_e$ has the exact RIR equal to 
$1/\|g_e\|_{L_\infty}$. According to the lemma, it is so if the following conditions hold: 
(a) $|e| < 1/\|g_e\|_{L_\infty}$; (b) $g_e$ has even number of ORHP poles; and (c) $g_e$ can be
marginally stabilized by a stable perturbation $\delta$ with $\|\delta\|_{H_\infty}= 1/\|g_e\|_{L_\infty}$. 
Numerical computations show that (a) holds for $e\in(-0.193,0.174)=:\mathbb{I}_1$ and that 
$g_e(s)\in {\cal G}_2^{\#}$, which implies that (b) is satisfied for $e\in[-0.94,1)=:\mathbb{I}_2$. 
Thus, $1/\|g_e\|_{L_\infty}$ gives the exact RIR for the interval $\mathbb{I}_1\cap\mathbb{I}_2=(-0.193,0.174)$
if (c) is satisfied. Condition (c) can be readily verified by the phase change rate condition 
\eqref{cond:exactRIRa} stated in Theorem~\ref{thm:exactRIR} and it indeed holds for any $e\in\mathbb{I}_1\cap\mathbb{I}_2$.  
It should be emphasized that Theorem~\ref{thm:exactRIR} is applicable to any rational functions,  
while its counterpart in~\cite{HIH:Automatica2021} holds only for a particular class of third-order transfer functions.

\begin{figure}[h]
\epsfig{file=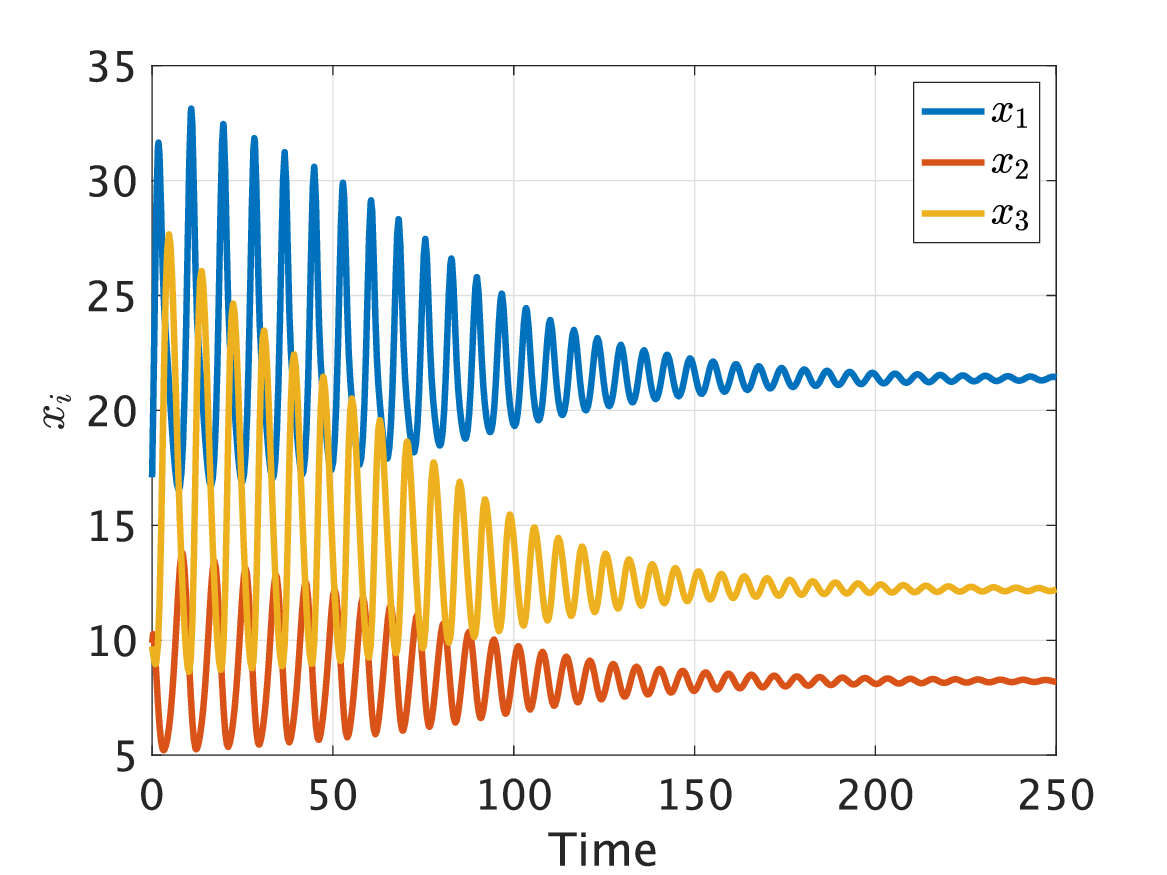, width=42.5mm}
\epsfig{file=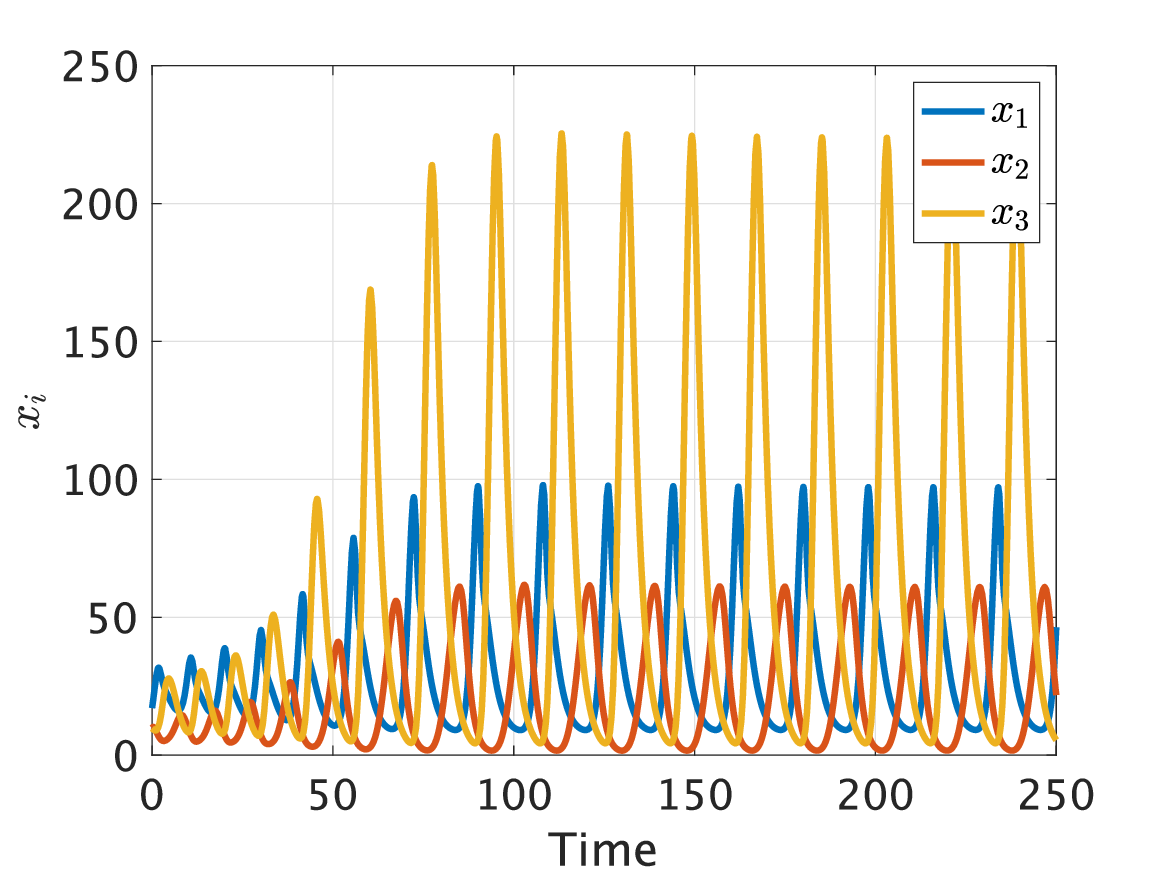, width=42.5mm}
\caption{Time-course simulations of the closed-loop systems. Left: $g_e(s)$ and $\delta(s)$ with $\epsilon=0.05$. 
Right: $g_e(s)$ and $\delta(s)$ with $\epsilon = -0.05$.}
\label{fig:3.2.2}
\end{figure}

For illustration purposes, let $e = 0.1$, and we construct a stabilizing perturbation $\delta(s)$ with $\|\delta\|_\infty = 1/|g_e(j\omega_p)| = 0.178$.
A stabilizing perturbation $\delta(s)$ is constructed using a first-order all-pass function and a high-pass filter that makes the DC gain $\delta(0)$ equal to $e$: 
\[
\delta(s) =(1 + \epsilon)  \frac{s + \gamma \xi}{s + \xi}  \cdot   
b\left(\frac{s-a}{s+a}\right), 
\]
where $a = 0.901$, $b = 0.178$, $\xi = 0.01$, and $\gamma = -e/(b(1+ \epsilon))$ with a non-negative constant $\epsilon$. 
The system is marginally stabilized when $\epsilon = 0$ and becomes asymptotically stable when the gain of $\delta(s)$ is slightly increased by making $\epsilon > 0$. 
The nonlinear repressilator models with $\epsilon = 0.05$ and $\epsilon = -0.05$ are simulated, and the results are shown in Fig.~\ref{fig:3.2.2} (left and right figures, 
respectively). Clearly, $\delta(s)$ with $\epsilon = -0.05$ is not able to stabilize $g_e(s)$ and the closed-loop system exhibits oscillatory 
behavior. On the other hand, $\delta(s)$ with $\epsilon = 0.05$ stabilizes $g_e(s)$ and the oscillatory behavior ceases to exist. 

\subsection{Discrete-time Strong Stabilization for Magnetic Levitation Systems}
\label{sec:meglev_sys}

A typical linearized model for the magnetic levitation system~\cite{Namerikawa2001} at an equilibrium is 
represented by 
\begin{align}
\label{glev}
g(s) = k/\left((-s^2+p^2)(\tau s+1)\right),
\end{align} 
where $k>0$, and the pair of poles at $\pm p$ is due to the mechanical aspect of the system 
while the stable pole at $-\tau^{-1}$ comes from the electrical part. 
A summary of the results in \cite{IFAC23_KHetal} on the minimum-norm continuous-time stabilization 
by a stable controller $c(s)$ is as follows: 
\begin{equation}
\label{lb_inf_g}
\frac{1}{\|g\|_{L_\infty}} = \frac{1}{|g(0)|} = \frac{p^2}{k}  
 < \inf_{c\in\IS(g)}\|c\|_{H_\infty} 
 \le \frac{1+p^2\tau^2}{\|g\|_{L_\infty}}  . 
\end{equation}
We can see that the upper bound approaches the lower bound when $\tau>0$ tends to zero, 
which is consistent with the case of reduced order model (i.e., $\tau=0$) where the infimum is exact. 

We consider  minimum-norm strong stabilization of~\eqref{glev} with $\tau=0$ in the digital control setting, 
where a reduced-order model
represented by 
\begin{align} \label{grlev} 
g_r(s) = k/(-s^2+p^2),
\end{align}
is used for simplicity to avoid the complicated formulae.  
This is reasonable in practice since $\tau^{-1}\gg p$ typically holds and hence 
we may neglect the factor $(\tau s+1)$ for control design purpose.

The first step is to derive the discretized plant model $g_d(z)$, where we assume that 
an ideal sampler and a synchronized zeroth-order hold with the sampling period $T>0$ are 
placed around the continuous-time plant $g_r(s)$. 
The discretized model with  one sample computational time delay ($1/z$) is given by 
\begin{align}
\label{eq:DiscML}
g_d(z) = \kappa(z + 1)/\left((z - \e^{-pT})(z - \e^{pT})z\right),
\end{align}
where $\kappa:=k(1-\e^{pT})(1-\e^{-pT})/(2p^2)$.  
Note that the static gain is preserved, i.e., $g_d(1) = g_r(0) = k/p^2$. 

Our theoretical results in the continuous-time setting can be applied to discrete-time systems by introducing 
a bilinear transformation $z\leftarrow(1+s)/(1-s)$. In particular, a continuous-time 
equivalent of $g_d(z)$, denoted by $g_{d,c}(s)$, can be derived as follows:
\begin{align*}
g_{d,c}(s)=k_c\frac{(s-1)^2}{(s-q)(s+q)(s+1)},  \quad \;\; 
q := \frac{1-\e^{-pT}}{1+\e^{-pT}}, 
\end{align*}
where $k_c := 2\kappa/((1+\e^{pT})(1+\e^{-pT}))$. 
Note that $k_c$ can also be expressed as $k_c = -kq^2/p^2$. 
Therefore, we have $g_{d,c}(0)= k/p^2 = g_r(0)$, which is also equal to $|g_d(1)|=\|g_d\|_{L_\infty}$. 
This is simply because the bilinear transformation preserves the gain of the 
corresponding frequency. 
It is also noticed that the unstable pole at $s=q$ is in the interval $(0,1)$, implying that it is located to the left of
the non-minimum-phase zero at $s=1$ and hence the PIP condition holds.  
 
One can readily verify that $g_{d,c}\in\mathcal{G}_1^0$ and $\theta_{g_{d,c}}'(0)<0$. To see this,
note that $A_{g_{d,c}}'(\omega)$ and $\theta_{g_{d,c}}'(\omega)$ are the real and imaginary parts
of $\frac{d}{d\omega}\log g_{d,c}(j\omega)$, respectively. They are found as follows
\begin{align*}
A_{g_{d,c}}'(\omega) = \frac{-\omega(\omega^2+2-q^2)}{(\omega^2+1)(\omega^2+q^2)}, \ \
\theta_{g_{d,c}}'(\omega) = \frac{-3}{\omega^2+1}.
\end{align*}
By these expressions, we have $A_{g_{d,c}}'(0)=0$, $A_{g_{d,c}}'(\omega)<0$ $\forall\omega>0$, and $\theta_{g_{d,c}}'(0)=-3$. 
Thus, by Theorem~\ref{thm:exactRIR}, we conclude that 
\begin{align*}
\inf_{c\in\mathbb{S}(g_{d,c})}\|c\|_{H_{\infty}} > 1/\|g_{d,c}\|_{L_{\infty}}
=1/|g_{d,c}(0)|=p^2/k
\end{align*}
since the necessary condition for $g_{d,c}$ to have the exact RIR is violated. This in turn implies that the norm of 
the minimum-norm strongly stabilizing controller for $g_d$ must also be larger than $p^2/k = 1/\|g_d\|_{L_\infty}$.  

To obtain an upper bound on $\inf_{c\in\mathbb{S}(g_{d,c})}\|c\|_{H_{\infty}}$ (which is also an upper bound
on the norm of the minimum-norm strongly stabilizing controller for $g_{d}$), let us apply the lead compensator of the form 
$\left(\frac{b s+1}{as+1}\right)^m:=f_{\ell}(s)$ to $g_{d,c}$, where 
$m\ge 1$ is an integer and $b> a>0$. The values of these parameters are to be chosen later. 
Let $\tilde{g}_{d,c}(s):=g_{d,c}(s) f_{\ell}(s)$. One can readily verify that
\begin{align*}
&\hspace{-0.3cm}\theta_{\tilde{g}_{d,c}}'(\omega) = \theta_{g_{d,c}}'(\omega) +m\cdot \frac{(a-b)ab\omega^2+(b-a)}{(b^2\omega^2+1)(a^2\omega^2+1)},\\
&\hspace{-0.3cm}A_{\tilde{g}_{d,c}}'(\omega) = A_{g_{d,c}}'(\omega)
+m\cdot\frac{\omega(b^2-a^2)}{(b^2\omega^2+1)(a^2\omega^2+1)}.
\end{align*}
Thus for $\theta_{\tilde{g}_{d,c}}'(0)=-3+m(b-a)>0$, we set $b=a+\3/m$, where
$\3:=3+\epsilon$, and $\epsilon>0$ is arbitrarily small. With this selection of $b$,  
$\theta_{\tilde{g}_{d,c}}'(0)>0$ and $A_{\tilde{g}_{d,c}}'(0)=0$ regardless of the value of $a$. 
Furthermore, with $b=a+\3/m$, 
it can be verified that $A_{\tilde{g}_{d,c}}'(\omega)\le 0$ when $\omega \to 0^+$ if and only if 
$a \le \frac{1}{\3}\left(1/q^2-(\3^2/m+1)/2\right)=:\bar{a}$. The condition also implies that $A_{\tilde{g}_{d,c}}'(\omega)\le 0$,
$\forall~\omega\ge 0$. Thus, $\tilde{g}_{d,c}$ belongs to $\mathcal{G}_1^0$ with positive PCR at the zero
frequency for any $a$ satisfying $0<a\le \bar{a}$. Note that
the upper bound $\bar{a}$ is positive if $m$ is sufficiently large, since $1/q^2$ is larger than $1$. 
Applying Theorem~\ref{thm:exactRIR}, we have
\begin{align*}
\inf_{c\in\mathbb{S}(\tilde{g}_{d,c})}\|c\|_{H_\infty} = 1/\|\tilde{g}_{d,c}\|_{L_\infty}=1/|\tilde{g}_{d,c}(0)|=p^2/k.
\end{align*}  
Suppose $c^*$ is a (minimum-norm) strongly stabilizing controller for $\tilde{g}_{d,c}$. Then $c^*f_{\ell}$ is a 
strongly stabilizing controller for $g_{d,c}$, and therefore $\|c^*f_{\ell}\|_{H_\infty} \approx (p^2/k)(1+\3/(ma))^m$ 
is an upper bound on $\inf_{c\in\mathbb{S}(g_{d,c})}\|c\|_{H_\infty}$. To minimize the upper bound, we shall choose
$a=\bar{a}$, which results in
\begin{align}
\hspace{-0.5cm}
1 < \inf_{c\in\mathbb{S}(g_{d,c})}\frac{\|c\|_{H_\infty}}{p^2/k} \le \left(1+\frac{2q^2\cdot\3^2}
{m(2-(\frac{\3^2}{m}+1)q^2)}\right)^m
\label{eq:mss_bound}
\end{align}
As a final remark, recall that $q:=(1-\e^{-pT})/(1+\e^{pT})$, which is a monotonically increasing function
of $pT$, and takes $0$ value when $pT=0$. As such, we see that the upper bound approaches the lower bound as $pT\to 0$; i.e.,
when the sampling is arbitrarily fast, we recover the continuous-time result, which is expected. Furthermore, for a given $pT$, one 
can try to minimize the upper bound over $m$ under the constraint $\bar{a}>0$; i.e., $m$ is a positive integer satisfying
$m>q^2\cdot\3^2/(2-q^2)$. 

\section{Proofs of main results}
\label{sec:proof_thm}
\subsection{Proof of Theorem~\ref{thm:Main}}
\label{subsec:ProofMainTheorem}

The arguments for proving both statements are similar; here we will focus on the derivation 
for statement (II), and then point to key differences that lead to statement (I).  

We start with proving the sufficiency part of statement (II). 
Let $g \in {\cal G}_n^\#$. 
Suppose $n=2$ and \eqref{eq:CRomega} holds. For the case $\sin(\theta_g(\omega_p)) = 0$, let 
$f$ be the constant function with real value $1/g(j\omega_p)$ and $L=gf$. As $g\in\mathcal{G}_2^{\#}$, we 
see that $L\in\mathcal{G}_2^{\#}$, and $|L(j\omega)|< L(j\omega_p) = 1$ for all $\omega\not = \pm \omega_p$. 
Furthermore, since $\theta_L'(\omega_p) = \theta_g'(\omega_p) + \theta_f'(\omega_p) = \theta_g'(\omega_p) > 0$, 
we also have $\left. \frac{d}{ds} L(s) \right|_{s=j\omega_p} \neq 0$. 
Thus, \eqref{eq:PeakGain} and condition (i) of Proposition~\ref{prop:OmegaMS2} hold with $\omega_c = \omega_p>0$. 
Lastly, since $|L(j\omega)|\le 1$ for all $\omega$,  we see that $0= \nu_o(0) = 2-2$ with $\theta_L'(\omega_p)>0$, 
and hence condition (ii-a) of Proposition~\ref{prop:OmegaMS2} also holds. 
By Proposition~\ref{prop:OmegaMS2}, we have that $f$ marginally stabilizes $g$. For the case 
$\sin(\theta_g(\omega_p)) \not= 0$, let 
\begin{align*}%
&\hspace{-0.7cm}
f(s) =\begin{cases}
 \frac{1}{\|g\|_{L_\infty}}\left(\frac{a-s}{a+s}\right) & \mbox{if $\theta_g(\omega_p)$ mod $2\pi\in (0, \pi)$}\\
 \frac{1}{\|g\|_{L_\infty}}\left(\frac{s-a}{s+a}\right) & \mbox{if $\theta_g(\omega_p)$ mod $2\pi\in (-\pi,0)$} 
\end{cases},
\end{align*}
where $a>0$ is chosen such that $\theta_f(\omega_p) + \theta_g(\omega_p) = 0\ (\mbox{mod $2\pi$})$. 
Specifically, $a = \omega_p/\tan(\theta_g(\omega_p)/2)$ when $\theta_g(\omega)$ (mod $2\pi$) $\in(0,\pi)$, and
$a=\omega_p\tan(-\theta_g(\omega_p)/2)$ when $\theta_g(\omega)$ (mod $2\pi$) $\in(-\pi,0)$. 
Since $f$ is an all-pass function satisfying
$g(j\omega_p)f(j\omega_p)=1$, the loop transfer function $L$ satisfies $|L(j\omega)|< L(j\omega_p) = 1$ 
for all $\omega\not = \pm \omega_p$. Furthermore, one can verify that for a given $\omega\not = 0$, 
$\theta_f'(j\omega)=-|\sin(\theta_f(\omega))/\omega|$; thus, we again have 
$\theta_L'(\omega_p) = \theta_g'(\omega_p) + \theta_f'(\omega_p) > 0$ by \eqref{eq:CRomega}. 
Therefore, the same arguments apply and we conclude that 
$f$ marginally stabilizes $g$ and the feedback system has a pair of single poles at $\pm j\omega_p$.
Note that, in both cases, the feedback system is $\omega_p$-marginally stable and 
the stabilizing (sub)system $f$ is stable and satisfies $\|f\|_{H_\infty} = 1/\|g\|_{L_\infty}$.

We note that the phase change rate condition~\eqref{eq:CRomega} (resp.~\eqref{eq:CR0} for 
case (I)) renders $\left. \frac{d}{ds} L(s) \right|_{s=j\omega_p} \neq 0$, which in turn implies 
that the closed-loop poles at $\pm j\omega_p$ (resp. at the origin for case (I)) must be simple. 

For the necessity part of statement (II), let $g\in {\cal G}_n^\#$ for which 
the peak gain of $g$ occurs at $\omega_p$. 
Let $f$ be the stable system that satisfies $\|f\|_{H_\infty} = 1/\|g\|_{L_\infty}$ and marginally
stabilizes $g$. Since the feedback system is marginally stable and the loop transfer function $L:=gf$ 
satisfies $\|L\|_{L_\infty}:=\|gf\|_{L_\infty} \le 1$, we must have $|L(j\omega)|=1$ for some frequency.
This can only occur when $f$ satisfies $1/\|g\|_{L_\infty} = |f(j\omega_p)| \ge |f(j\omega)|$ for 
any $\omega\not=\pm\omega_p$, which in turn implies that 
$ 1 = |L(j\omega_p)| > |L(j\omega)|\quad \forall \ \omega\not=\pm\omega_p$.  
Since $f$ is stable, we have $L\in{\cal G}_n^\#$ with peak gain occurring at $\omega_p$ and equal to 1. 
This further implies that the feedback system must be $\omega_p$-marginally stable. 
By Proposition~\ref{prop:OmegaMS2}, $L$ must satisfy condition (ii-a) or condition (ii-b). %
Note that $\|L\|_{L_\infty} \le 1$ implies that $\nu_o(0)=0$ and condition (ii-b) 
can never be satisfied. Thus, condition (ii-a) must hold and we conclude that $n=2$ as 
the corresponding $\omega_c$ is equal to $\omega_p$, which is strictly larger than 0. Moreover, 
the necessary condition $\theta_L'(\omega_p) > 0$
implies that $\theta_g'(\omega_p) > - \theta_f'(\omega_p)$. Inequality (\ref{eq:CRomega}) emerges 
as one finds the infimum of $-\theta'_f(\omega_p)$ over the set of all suitable $f$'s. 
This results in the optimization problem we stated in~\eqref{eq:GPCR}, and the claim 
thereafter gives condition~\eqref{eq:CRomega}.

Regarding statement (I), we note that the statement deals with the special case where $\omega_p = 0$; hence
its proof is almost identical to that of statement (II). 
For sufficiency, notice that since $g(j0)$ is real, the arguments for the case where 
$\sin(\theta_g(\omega_p))=0$ apply. The marginally stabilizing $f$ is equal to $1/g(j0)$. The remaining
arguments are identical, except when we verify condition (ii-a) of Proposition~\ref{prop:OmegaMS2}, we have
the corresponding $\omega_c=0$ and $n=1$. For necessity, again, the same arguments for proving the necessity part 
of statement (II) apply. As such, condition (ii-a) of Lemma~\ref{lemma:MS} implies that $n=1$, and
$\theta_g'(0)$ must satisfy $ \theta_g'(0) > - \theta_f'(0)$. Inequality~\eqref{eq:CR0} 
emerges as the infimum of the right hand side is equal to 0, as we have stated and has been shown in 
Section~\ref{sec:SupCR}.

\subsection{Proof of Theorem~\ref{thm:exactRIR}}
\label{subsec:ProofExactRIR}
{ 
To prove statement (I), we need the following lemma, whose proof is given in Appendix~\ref{appendix:PS}.
\begin{lemma} \label{lemma:PS}
Given $\omega_c\ge 0$, an integer $n\ge 1$, and a transfer function $L\in\mathcal{G}_n$,  
consider the positive feedback system with loop transfer function $L$ satisfying the following condition
\begin{align}
\begin{split}
&1=|L(j\omega_c)|=\|L\|_{L_\infty}, \\
&\hspace{2cm} |L(j\omega)|<|L(j\omega_c)|, \forall \omega\not=\pm\omega_c.
\end{split}\label{ineq:PCR2}
\end{align}
If the closed-loop system has all its poles in the CLHP, then $\theta_L'(\omega_c)\ge 0$. 
\end{lemma}

Now suppose $g\in\mathcal{G}_n^0$ has the exact RIR equal to $1/\|g\|_{L_\infty}$. 
This implies that one can find a 
stable $f$, $\|f\|_{H_\infty}=1/\|g\|_{L_\infty}$, 
such that the closed-loop system, with loop transfer function $L:=gf$, has all its poles in the CLHP. Note 
that $\|L\|_{L_\infty}\le 1$; therefore
$L$ must satisfy~\eqref{ineq:PCR2} with $\omega_c=0$, whereby $\theta_L'(0)=\theta_g'(0)+\theta_f'(0) \ge 0$. This
implies $\theta_g'(0)\ge -\theta_f'(0)$, and hence $\theta_g'(0)\ge 0$ as the infimum of the right hand side is equal 
to 0, which is stated in the proof of Theorem~\ref{thm:Main} and has been shown in Section~\ref{sec:SupCR}. The 
arguments for $g\in\mathcal{G}_n^{\#}$ are identical, except that in this case we have $\omega_c:=\omega_p >0$, 
$\theta_g'(\omega_p)\ge -\theta_f'(\omega_p)$, and the infimum of the right hand side is now 
$|\sin\left(\theta_g(\omega_p)\right)/\omega_p|$. 

To see statement (II), note that by Theorem~\ref{thm:Main}, condition~\eqref{eq:CRomega} 
implies $g$ can be marginally stabilized with a single pole at $0$ (for $n=1$) or $\pm j\omega_p$ (for $n=2$) 
by a stable system $f$ with $\|f\|_{H_\infty} = 1/\|g\|_{L_\infty}$. Further, by Proposition~\ref{prop:lbub}, 
an arbitrarily small perturbation of $f$ can exponentially stabilize $g$. 
Thus, by definition, $\rho_*(g) = 1/\|g\|_{L_\infty}$. 

To prove statement (III), consider a stable $f$ and the feedback 
system with loop transfer $k\cdot L$, where $k\in(0,1]$ 
and $L:=gf$. Suppose that, when $k=1$ all poles of the closed-loop system are in the CLHP. 
Since $L$ has an odd number of unstable poles, at least one unstable pole must be on the positive 
real axis, and there must be a $k\in(0,1]$ such that the closed-loop system has at least one pole 
at the origin. That is to say,  
$1 = k\cdot g(0)f(0)$ for some $k\in(0,1]$. 
This implies $|f(0)| = 1/(k\cdot |g(0)|) \ge 1/|g(0)| > 1/\|g\|_{L_\infty}$. The last inequality 
follows from the fact that $|g(0)|<|g(j\omega_p)|:=\|g\|_{L_\infty}$ (for some $\omega_p>0$) since $g\in\mathcal{G}_n^{\#}$. This shows 
that, any stable $f$ which renders all closed-loop poles in the CLHP must satisfy $\|f\|_{H_\infty}>1/\|g\|_{L_\infty}$.
Hence, $\rho_*(g)>1/\|g\|_{L_\infty}$.
}
\subsection{Proof of {  Proposition}~\ref{thm:AP}}
\label{subsec:ProofAP}
{  To prove Proposition~\ref{thm:AP}, we require the results stated in the following technical lemmas. 
The proofs of these lemmas can be found in the Appendix. 
\begin{lemma} \label{lem:negEq}
For any $f\in\RHinf$, we have  
\begin{equation} \label{eq:MFsym}
M_f'(\omega) =  M_{-f}'(\omega) = M_f'(-\omega). 
\end{equation}
\end{lemma}
\begin{lemma} \label{lem:ineq_sin} 
Let $\theta_i$ belong to $[-\pi, 0]$, $i=1,\cdots,n$. Then 
\begin{align}
\label{ineq:sum_sin}
\sin(\theta_1)+\cdots+\sin(\theta_n) \le -|\sin(\theta_1+\cdots+\theta_n)|.
\end{align}
\end{lemma}

The next result provides an upper bound for the $\sigma$-gain change rate as in~\eqref{eq:S_GainPhaseCR} 
of any all-pass function that has real poles. Notice that the bound is tight for zeroth-order or first-order all-pass functions. 

\begin{lemma}  \label{thm:rateReal} 
Let $H(s)$ be a stable $n^{\rm th}$-order all-pass function with real poles (or no pole when $n=0$). 
Then for a given $\omega_p\not = 0$,
\begin{align}
M_H'(\omega_p) \le  -\left|\sin(\theta_{H}(\omega_p))/\omega_p\right| . 
\label{eq:RateChangeReal_p1}
\end{align}
Moreover, when $n=0$ and $n=1$ the equality holds for all $\omega_p\not= 0$.\footnote{ 
When $n=0$, clearly $M_H'(\omega_p)=0$ for all $\omega_p$, including $\omega_p=0$.} 
\end{lemma}

By Lemma~\ref{lemma:GPrates}, we have the same upper bound for the phase change rate 
$\theta_H'(\omega_p)$; i.e., 
$\theta_H'(\omega_p) \le  -\left|\sin(\theta_{H}(\omega_p))/\omega_p\right|$
for any $\omega_p\not= 0$. Our next result shows that the $\sigma$-gain change rate (and hence the phase change rate) of 
a second-order all-pass function with complex poles is upper bounded by that of a corresponding second-order all-pass 
function with real poles. 

\begin{lemma}  \label{thm:c<r} 
Let $H_r(s)$ and $H_c(s)$ be rational second-order stable all-pass functions with two real poles 
and a pair of complex conjugate poles, respectively.
For given $\omega_p\not = 0$, if $\theta_{H_r}(\omega_p) = \theta_{H_c}(\omega_p)$, then 
\begin{align}
\label{ineq:c<r}
M_{H_c}'(\omega_p)  < M_{H_r}'(\omega_p).
\end{align}
\end{lemma} 

Now we are ready to prove Proposition~\ref{thm:AP}. 
Let $f$ be an $n^{\rm th}$-order real-rational all-pass function with unit norm. Then
$f$ can be factorized as $\pm f_1f_2\cdots f_m$, where $f_i$, $i=1,\cdots,m$ is either 
a first-order all-pass function of the form $\frac{a-s}{a+s}$, where $a>0$, 
or a second-order one of the form $\frac{a-bs+s^2}{a+bs+s^2}$, where $a, b>0$ are such that
the poles are a pair of complex conjugates. This follows from results on 
finite Blaschke products~\cite{Garcia_etal:BlaschkeProduct}. 

First consider the case where $\omega_p\not=0$. By Lemma~\ref{thm:c<r}, 
any second-order factor $f_c$ of $f$ can be replaced by the product of two first-order all-pass functions
$f_r:=f_{r1}f_{r2}$, such that $\theta_{f_c}(\omega_p)=\theta_{f_r}(\omega_p)$ and 
$M_{f_c}'(\omega_p)< M_{f_r}'(\omega_p)$. As a result, the supremum of~\eqref{eq:Problem_AP} can be 
found by searching over the subset of $\AP_{\omega_p,\theta_{p}}$ 
containing real-rational all-pass functions which are constant functions or can be factorized
as products of first-order factors. Furthermore, among these all-pass functions,
Lemma~\ref{thm:rateReal} shows that 
\[
M_f'(\omega_p)
\le -\left|\frac{\sin(\theta_{f}(\omega_p))}{\omega_p}\right|
= -\left|\frac{\sin(\theta_{p})}{\omega_p}\right| \le 0.
\]
When $\theta_{p}\not\in\{0,\pi\}$, the supremum is attained by first order all-pass function in the form of 
$\frac{a-s}{a+s}$ or $\frac{s-a}{a+s}$. The value of $a>0$ is chosen such that 
$\sin(\theta_{p}) = \frac{-2a\omega_p}{a^2+\omega_p^2}<0$, or  
$\sin(\theta_{p}) = \frac{2a\omega_p}{a^2+\omega_p^2}>0$.
One can readily verify that such an $a$ always exists. When $\theta_{p}\in\{0,\pi\}$, 
it is clear that $f(s)=1$ or $f(s)=-1$ would satisfy the angular constraint and attain the supremum of the
phase change rate, which is equal to $0$.  

For the case where $\omega_p=0$, first we note that, since we consider real rational function $f$, $f(j0)$ 
is necessarily real and thus $\theta_f(0)$ can only assume value $0$ or $\pi$ (mod $2\pi$). 
Since $M_f'(\cdot) = M_{-f}'(\cdot)$ by Lemma~\ref{lem:negEq}, without loss of generality 
we can consider only all-pass functions $f$ such that $f(j0)>0$; i.e., the corresponding
$\theta_p$ is set to 0. The factorization of $f$ implies that 
$M_f'(\cdot) = \sum_{i=1}^m M_{f_i}'(\cdot)$. Note that for a first-order factor of the form
$\frac{a-s}{a+s}$, the $\sigma$-gain change rate at $0$ is equal to $-2/a$, while for 
a second-order factor of the form $\frac{a-bs+s^2}{a+bs+s^2}$, the change rate is $-2b/a$. 
Therefore, 
one concludes that $M_f'(0)$ is strictly less than 0 if the order of $f$ is larger than or equal to $1$. 
Hence, the supremum of $M_f'(0)$ ($=\theta_f'(0)$) is equal to 0, obtained by the zeroth-order all-pass
functions $f(s)=1$ or $f(s)=-1$. 
\subsection{Proof of Proposition~\ref{prop:FR=AP}}
\label{subsec:ProofSupHinf}

To prove Proposition~\ref{prop:FR=AP}, we need the following lemma which provides an upper bound for the 
phase change rate of any minimum-phase stable function at the peak-gain frequency. 
\begin{lemma}  \label{thm:bound} 
Given minimum-phase $f\in\RHinf$, 
suppose $\omega_p \not = 0$, $\theta_f(\omega_p) \in (-\pi, \pi]$, and $|f(j\omega_p)|=\|f\|_{H_{\infty}}$. Then
\begin{align*}
\theta_f^\prime(\omega_p) \leq - \left|\theta_f(\omega_p)/\omega_p\right|.
\end{align*}
Moreover, if $\omega_p=0$, then $\theta_f'(\omega_p) \le 0$.
\end{lemma}

The proof of Lemma~\ref{thm:bound} can be found in the Appendix. To facilitate the development, 
let us also define
\begin{align*}
\mathcal{O}_{\omega_p, \theta_p} := \{f & \in \RHinf : f \text{ is minimum-phase, }\\
 &|f(j\omega_p)| = \|f\|_{H_{\infty}}, \text{ and } \theta_f(\omega_p) = \theta_p\}.
\end{align*}

First consider the case where $\omega_p\not = 0$ and $\theta_p\in(-\pi,\pi]$. 
Since every $f \in \RHinf$ admits an inner-outer factorization $f = f_if_o$ with $f_i \in \RHinf$ 
being all-pass and $f_o \in \RHinf$ being minimum-phase, it follows that ${  |f_o(j\omega_p)|} = \|f_o\|_{H_\infty}$ and
  \begin{align*}
  &\hspace{-0.7cm} \sup_{f \in \F_{\omega_p, \theta_p}} \theta_f^\prime (\omega_p) 
     = \sup_{f = f_if_o \in \F_{\omega_p, \theta_p}} \left(  \theta_{f_i}^\prime (\omega_p) +
        \theta_{f_o}^\prime (\omega_p) \right) \\
    &\hspace{-0.7cm} \le \sup_{\theta \in (-\pi, \pi]} \left( \sup_{f_i \in \AP_{\omega_p, \theta}}  \theta_{f_i}^\prime
      (\omega_p) + \sup_{f_o \in {  \mathcal{O}_{\omega_p, \theta_p - \theta}} } \theta_{f_o}^\prime (\omega_p) \right) \\
    &\hspace{-0.7cm} \le \sup_{\theta \in (-\pi, \pi]} \left( -\frac{|\sin(\theta)|}{|\omega_p|} -
      \frac{|\theta_p - \theta|}{|\omega_p|} \right) 
    \le -\left|\frac{\sin(\theta_p)}{\omega_p}\right|,
  \end{align*}
where the second inequality follows from Proposition~\ref{thm:AP} and Lemma~\ref{thm:bound}, and the last inequality
follows from the fact that $|b - a| + |\sin(a)| - |\sin(b)| \geq 0$ for all $a, b \in (-\pi, \pi]$.
Lastly, {  since $-\left|\sin(\theta_p)/\omega_p\right| = 
\sup_{f \in \AP_{\omega_p, \theta_p}} \theta_f^\prime (\omega_p)  \le \sup_{f \in \F_{\omega_p, \theta_p}} \theta_f^\prime (\omega_p)$, 
we conclude that the supremum of $\theta'(\omega_p)$ over $\F_{\omega_p, \theta_p}$ is the same as that over $\AP_{\omega_p, \theta_p}$.}

Now consider the case where $\omega_p=0$ and $\theta_p\in\{0,\pi\}$. The reason why the value of 
$\theta_p$ is restricted to $0$ or $\pi$ is due to $f$ being real rational and therefore 
$\theta_f(\omega_p)\in\{0,\pi\}$. The rest of the derivation follows the same arguments. By the inner-outer 
  factorization of $f$, we have
  \begin{align*}
    &\sup_{f \in \F_{0, \theta_p}} \theta_f^\prime (0) 
     = \sup_{f = f_if_o \in \F_{0, \theta_p}} \left(  \theta_{f_i}^\prime (0) +
        \theta_{f_o}^\prime (0) \right) \\
    & \le \sup_{\theta \in \{0,\pi\}} \left( \sup_{f_i \in \AP_{0, \theta}}  \theta_{f_i}^\prime
      (0) + \sup_{f_o \in \mathcal{RF}_{0, \theta_p - \theta}} \theta_{f_o}^\prime (0) \right) \\
    & \le \sup_{\theta \in \{0,\pi\}} \sup_{f_i \in \AP_{0, \theta}}  \theta_{f_i}^\prime(0)=0.
  \end{align*}
The second inequality follows from Lemma~\ref{thm:bound}, where it is shown that $\theta_{f_o}'(\omega_p) \le 0$ 
for any minimum-phase $f_o\in\RHinf$. The last equality follows from Proposition~\ref{thm:AP}, which concludes the proof.

\section{Conclusion}
\label{sec:Concl}

This paper examined the exact RIR condition motivated {  by} robust instability analysis 
against stable perturbations and minimum-norm strong stabilization. 
We have shown that the problem of finding the exact RIR may be turned into {  the} problem of 
maximizing the phase change rate at the peak frequency with a phase constraint. 
It has been proven that the supremum is attained by a constant or a first-order all-pass function, which 
yields conditions for which we can {  find} the exact RIR in terms of the phase change rate. 
Two practical applications have been provided to demonstrate the
effectiveness of our theoretical results in practice.

\appendix

\subsection{Proof of Lemma~\ref{lem:CR_int_relations}}
\label{appendix:CRint}
Given a minimum-phase function $f\in\RHinf$, let $h(s) := \frac{d}{ds} \log f(s)$ $= \frac{f^\prime(s)}{f(s)}$. Note that
$h\in\RHinf$ and is strictly proper, and $h(j\omega) =  \theta_f^\prime(j\omega) - j A_f^\prime(j\omega)$.

Now let $k(s) := s \cdot h(s) \in\RHinf$. Then, we have 
\[
k(j\omega) = j\omega h(j\omega) = \omega A_f^\prime(j\omega) + j \omega \theta_f^\prime(j\omega).
\]
{  Following the derivation in~\cite[Sec. 3]{Douce2007} gives the integral formula for $\theta_f^\prime(j\omega_p)$ stated in~\eqref{eq:int_f_1}.}
Now rewrite the right-hand-side of \eqref{eq:int_f_1} as
\begin{align*}
\frac{2}{\pi} \lim_{\epsilon \to 0} 
\left[ \int_0^{\omega_p - \epsilon} \frac{\omega A_f^\prime(j\omega)}{\omega^2 - \omega_p^2} \,
d\omega + \int_{\omega_p + \epsilon}^\infty \frac{\omega A_f^\prime(j\omega)}{\omega^2 - \omega_p^2} \, d\omega \right].
\end{align*}
Applying integration by parts to the two terms in the parentheses yields
\begin{align*}
&\hspace{-0.5cm}\left.\frac{\omega A_f(j\omega)}{\omega^2-\omega_p^2}\right|_0^{\omega_p -\epsilon} 
+\left.\frac{\omega A_f(j\omega)}{\omega^2-\omega_p^2}\right|_{\omega_p +\epsilon}^\infty + \\ 
&\hspace{-0.75cm} \int_0^{\omega_p-\epsilon} \frac{(\omega^2+\omega_p^2) A_f(j\omega)}{(\omega^2-\omega_p^2)^2}d\omega 
  +\int_{\omega_p+\epsilon}^\infty \frac{(\omega^2+\omega_p^2) A_f(j\omega)}{(\omega^2-\omega_p^2)^2}d\omega . 
\end{align*}
Since 
\begin{align*}
\lim_{\omega \to 0}\frac{\omega A_f(j\omega)}{\omega^2-\omega_p^2} = \lim_{\omega \to  \infty}\frac{\omega A_f(j\omega)}{\omega^2-\omega_p^2} = 0
\end{align*}
holds, the boundary terms become
\begin{align*}
&\frac{(\omega_p-\epsilon) A_f(j(\omega_p-\epsilon))}{(\omega_p-\epsilon)^2-\omega_p^2} - \frac{(\omega_p+\epsilon) A_f(j(\omega_p+\epsilon))}{(\omega_p+\epsilon)^2-\omega_p^2}  \\
=&\left(\frac{2\omega_p^2-\epsilon^2}{\epsilon^2-4\omega_p^2}\right)\frac{A_f(j(\omega_p-\epsilon))+A_f(j(\omega_p+\epsilon))}{\epsilon}\\
&+\frac{\omega_p(A_f(j(\omega_p+\epsilon))-A_f(j(\omega_p-\epsilon)))}{\epsilon^2-4\omega_p^2} . 
\end{align*}
Note that $A_f(j\omega_p)=0$, and thus by L'H\^{o}pital's rule we have 
\begin{align*}
 &\lim_{\epsilon\to 0}\frac{A_f(j(\omega_p-\epsilon))+A_f(j(\omega_p+\epsilon))}{\epsilon} \\
=&\lim_{\epsilon\to 0} -A_f'(j(\omega_p-\epsilon))+A_f'(j(\omega_p+\epsilon)) = 0 , 
\end{align*}
which in turn implies that the boundary terms approach zero as $\epsilon\to 0$.  
Thus we have 
\begin{align*}
  & \hspace{-0.7cm} \int_0^{\infty} \frac{\omega A_f'(j\omega)}{\omega^2-\omega_p^2}d\omega 
= \lim_{\epsilon \to 0}\left( \int_0^{\omega_p-\epsilon} \frac{(\omega^2+\omega_p^2) A_f(j\omega)}{(\omega^2-\omega_p^2)^2}d\omega \right. + \\
& \hspace{-0.7cm} \left. \int_{\omega_p+\epsilon}^\infty \frac{(\omega^2+\omega_p^2) A_f(j\omega)}{(\omega^2-\omega_p^2)^2}d\omega  \right)
= \int_0^{\infty}  A_f(j\omega) \frac{(\omega^2+\omega_p^2)}{(\omega^2-\omega_p^2)^2}d\omega 
\end{align*}
as claimed in the second part of the lemma. 
\subsection{Proof of Lemma~\ref{lemma:MS}}
\label{appendix:MS}
The idea for a proof of (i) is based on the Nyquist stability criterion, 
where a positive feedback system with unstable loop transfer function is stable if and 
only if the number of counter-clockwise encirclements by the Nyquist plot
about $1+j0$ is equal to the number of open-loop poles in the ORHP, ``$n$'' in this case.  
The proof is trivial by considering the Nyquist plot of $L(j\omega + \epsilon)$ 
with a sufficiently small positive number $\epsilon > 0$ instead of the original 
Nyquist plot of $L(j\omega)$ so that any pole of the closed-loop system on the imaginary axis
are placed outside of the Nyquist contour. Noting that $L$ is a strictly proper $\RLinf$ 
function, the equivalence of (i) and (ii) follows from continuity and 
boundedness of  $L(j\omega + \epsilon)$ in addition to the property
$L(j\omega + \epsilon)\rightarrow 0$ as $\omega\rightarrow\pm\infty$. 

It is clear that conditions (iii-a) and (iii-b) are necessary for marginal stability 
since they mean existence of simple roots of $1+L(j\omega)=0$ on the imaginary axis. 
Conversely, condition (i) or (ii) implies that all the poles are in the closed left half plane,
and hence additional conditions (iii-a) and (iii-b) are sufficient  
for marginal stability.  

\vspace{-0.25cm} 

\subsection{Proof of Proposition~\ref{prop:OmegaMS2}}
\label{appendix:OmegaMS2} 

First note that the phase change rate at $\omega = \omega_c$ cannot 
be equal to zero. The reason is as follows.
Suppose $\theta_L'(\omega_c) = 0$, then together with \eqref{eq:PeakGain} this implies $L'(j\omega_c) = 0$, 
i.e., there are at least two feedback poles at $j\omega_c$. 
Therefore, $\theta_L'(\omega_c)$ cannot be 0.
As such, the strict inequality conditions \eqref{eq:nonnegativeCR} and \eqref{eq:nonpositiveCR} can be 
equivalently replaced by $\theta_L'(\omega_c) \geq 0$ and $\theta_L'(\omega_c) \leq 0$, respectively. 
They will be treated {  as such} when we prove the necessity of these two conditions. 

{\it Necessity}: 
Condition (i) is necessary from the requirement on the closed-loop pole location 
on the imaginary axis for $\omega_c$-marginal stability.  
The necessity of condition (ii-a) or (ii-b) is derived from condition (ii) in 
Lemma~\ref{lemma:MS}, which is $\nu_o(\epsilon)=n$ for sufficiently small $\epsilon>0$.

When $\omega_c > 0$, condition (i) implies that there are two 
segments of the Nyquist plot of $L(j\omega)$ passing through 
the critical point $1+j0$, which corresponds to the two different frequencies $\pm \omega_c$. 
Moreover, we see that the two curves go across the real axis in the same 
direction
since the Nyquist plot is symmetric about the real axis, i.e., 
$L(j\omega)$ and $L(-j\omega)$ are complex conjugate to each other for all $\omega\in\IR$.  
This means that the possible increment or decrement of $\nu_o(\epsilon)$
for $L(j\omega + \epsilon)$ is two or zero when $\epsilon$ is slightly perturbed away 
from zero.
Hence, we have the following three possibilities: 
$\nu_o(0)=n-2$, $n$, and $n+2$.
Note that the first requirements of conditions (ii-a) and (ii-b) correspond to cases 
$\nu_o(0)=n-2$ and $\nu_o(0)=n$, respectively.

Let us first consider the case $\nu_o(0)=n-2$,
where the necessity of the second condition of (ii-a) can be shown 
by contradiction. Suppose the phase change rate is negative. 
Since $\theta'_L(\omega_c)=M'_L(\omega_c)<0$,  %
the Nyquist plot of $L(j\omega + \epsilon)$ 
with small $\epsilon>0$ crosses the real axis to the left of the critical point $1+j0$ 
in the downward direction as seen in the second row of the top figure of Fig.~\ref{fig:ex}.  
This implies that $\nu_+(\epsilon)$ and $\nu_-(\epsilon)$ are constant for 
sufficiently small $\epsilon\geq0$, 
and hence $\nu_o(\epsilon)= n-2$ is preserved for small $\epsilon>0$. 
This means violation of condition (ii) in Lemma~\ref{lemma:MS}, implying that
the feedback system is not marginally stable. By contradiction, 
the necessity of the second condition (ii-a) is proved. 

The idea for showing the necessity of condition (ii-b) for the case $\nu_o(0)=n$
is the same as that of condition (ii-a), 
i.e., by contradiction.  We assume that the phase change rate is positive, 
which yields that the Nyquist plot of $L(j\omega + \epsilon)$ with 
small $\epsilon>0$ passes through the real axis on the right of the critical point 
$1+j0$ in the upward direction as seen in the first row of the top figure of Fig.~\ref{fig:ex}.   
This implies that $\nu_+(\epsilon)$ increments by $2$ and that $\nu_-(\epsilon)$ 
does not change when $\epsilon$ is perturbed positively away from zero. Therefore,
we have $\nu_o(\epsilon)=n+2$ for sufficiently small $\eps>0$, contradicting 
condition (ii) in Lemma~\ref{lemma:MS}. Thus,
the necessity of the second condition (ii-b) is proved. 

It should be noticed that the case $\nu_o(0)=n+2$ never happens, because 
the change of the number ($\nu_o(0)\neq\nu_o(\epsilon)$) with small
perturbation $\epsilon>0$ occurs only if the $\sigma$-gain change rate 
$M_L'(j\omega_c)$
is nonnegative as seen in the first row of the top figure of Fig.~\ref{fig:ex}, which always gives the 
increase of the number ($\nu_o(0)<\nu_o(\epsilon)=n$)
due to the non-negativity of the phase change rate $\theta_L'(j\omega_c)$.

Finally, the proof for the case $\omega_c = 0$ is almost the 
same as that for $\omega_c > 0$. The only difference is the number of 
the Nyquist plot segments crossing the real axis at the critical point $1+j0$,
which is one instead of two when $\omega_c = 0$. Other than this difference, 
the arguments above are still valid. 

{\it Sufficiency}: 
The proof of the sufficiency is similar to that of the necessity,  and hence 
we only outline sufficiency of condition (ii-a). 
When the phase (or $\sigma$-gain) change rate is positive, the Nyquist plot of $L(j\omega_c + \epsilon)$ 
passes through the real axis on the right of the critical point $1+j0$ 
in the upward direction as shown in the first row of the top figure of Fig.~\ref{fig:ex}. 
This implies that the number of crossing points on the real semi-interval $(1, +\infty)$ 
from the negative imaginary region to the positive imaginary region increases by 
one (resp. two) for $\omega_c=0$ (resp. $\omega_c>0$). 
Hence we have $\nu_o(\epsilon)=n$ for sufficiently small $\epsilon>0$, which 
guarantees the marginal stability due to Lemma~\ref{lemma:MS}.

\vspace{-0.25cm} 

\subsection{Proof of Lemma~\ref{lemma:PS}}
\label{appendix:PS}
Define $\nu_o(\cdot)$ as in Proposition~\ref{prop:OmegaMS2}. Clearly, $\nu_o(0)=0$ due to the fact that
$|L(j\omega)|\le 1$ for all $\omega$. Suppose $\theta_L'(\omega_c) <0 $. By the arguments stated in 
Proposition~\ref{prop:OmegaMS2}, this would imply
that $\nu_o(\epsilon) = \nu_o(0)=0$ for sufficiently small positive $\epsilon$, which in turn implies that the feedback
system does not have all its poles in the CLHP by Lemma~\ref{lemma:MS}. Note that by Lemma~\ref{lemma:MS},
$\nu_o(\epsilon)=n$ is needed for having all closed-loop poles in the CLHP. 

\vspace{-0.25cm} 
\subsection{Proof of Lemma~\ref{lem:negEq}}
To see the first equality, simply note that 
$|f(\sigma+j\omega)| = |-f(\sigma+j\omega)|$. 
To see the second inequality, note that  
$f(\sigma-j\omega) = \bar{f}(\sigma+j\omega)$, 
where $\bar{f}(\cdot)$ denotes the complex conjugate of $f(\cdot)$. This is because
$f$ is a real rational function. Hence, we again have 
{  $\ln|f(\sigma-j\omega)|=\ln|f(\sigma+j\omega)|$}, and therefore $M_f'(\omega) = M_f'(-\omega)$. 

\vspace{-0.25cm} 
{ 
\subsection{Proof of Lemma~\ref{lem:ineq_sin}}
Let $k$ be a non-negative integer such that
$\theta_1+\cdots+\theta_n+k\pi \in [-\pi,0]$. Note that~\eqref{ineq:sum_sin} is equivalent to
\begin{align}
\sin\theta_1+\cdots+\sin\theta_n  \le \sin(\theta_1+\cdots+\theta_n+ k\pi). 
\label{ineq:sum_sin_eq}
\end{align}
To see this, notice that the right-hand side is equal to $\sin(\theta_1+\cdots+\theta_n)$
if $k$ is even, and $-\sin(\theta_1+\cdots+\theta_n)$ if $k$ is odd. Since the right-hand side
is always non-positive given the range of $\theta_1+\cdots+\theta_n+k\pi$, the right-hand side
of~\eqref{ineq:sum_sin_eq} is equivalent to $-|\sin(\theta_1+\cdots+\theta_n)|$. Given this, 
we will proceed to prove~\eqref{ineq:sum_sin_eq}.

The claim can be proven by induction. The claim is obviously true when $n=1$. Suppose it is true 
for $i=1,\cdots,n-1$. With $n$ terms, we have 
$\sum_{i=1}^n\sin\theta_i \le \sin \hat{\theta} + \sin\theta_n$, 
where $\hat{\theta}=\theta_1+\cdots+\theta_{n-1} + m\pi$ for some $m$ such that $\hat{\theta}\in [-\pi, 0)$. 
The rest of the proof 
boils down to {  analyzing} three possible scenarios: both $\hat{\theta}$ and $\theta_n$ belong to $[-\pi/2,0]$, 
both $\hat{\theta}$ and $\theta_n$ belong to $[-\pi,-\pi/2)$, and lastly, one is in $[-\pi/2,0]$ and another is in $[-\pi,-\pi/2)$. 

Suppose both $\hat{\theta}$ and $\theta_n$ belong to $[-\pi/2,0]$. Then we have $\hat{\theta}+\theta_n \in [-\pi, 0]$, and 
$\sin \hat{\theta}+\sin\theta_n  \le \sin\hat{\theta}\cos\theta_n + \sin\theta_n\cos\hat{\theta} = \sin(\hat{\theta}+\theta_n)$. The inequality holds
because $\cos\theta_n, \ \cos\hat{\theta} \in[0,1]$ and $\sin\theta_n, \ \sin\hat{\theta} \in [-1, 0]$. Thus we conclude
$\sum_{i=1}^n\sin\theta_i \le \sin(\sum_{i=1}^n \theta_i + k\pi)$, where $k:=m$ and $\sum_{i=1}^n \theta_i + k\pi\in[-\pi,0]$. 

Now suppose both $\hat{\theta}$ and $\theta_n$ belong to $[-\pi, -\pi/2)$. 
We have $\hat{\theta}+\theta_n \in [-2\pi, -\pi)$ 
and $\cos\theta_n, \cos\hat{\theta} \in[-1,0)$. This implies $\sin \hat{\theta}+\sin\theta_n \le \sin\hat{\theta}(-\cos\theta_n) + \sin\theta_n(-\cos\hat{\theta}) = -\sin(\hat{\theta}+\theta_n) = \sin(\hat{\theta}+\theta_n + \pi)$. 
Thus we again conclude 
$\sum_{i=1}^n\sin\theta_i \le \sin(\sum_{i=1}^n \theta_i + k\pi)$, where $k:=m+1$ and 
$\sum_{i=1}^n \theta_i + k\pi\in[-\pi,0]$.

Lastly, suppose without loss of generality that $\theta_n\in[-\pi/2,0]$ and $\hat{\theta}\in[-\pi,-\pi/2)$. 
This time,
$\hat{\theta}+\theta_n \in [-3\pi/2, -\pi/2)$. Depending on $\hat{\theta}+\theta_n \in [-\pi, -\pi/2)$ or $\hat{\theta}+\theta_n \in [-3\pi/2, -\pi)$,
we apply arguments similar to those in the previous two paragraphs to conclude inequality~\eqref{ineq:sum_sin_eq}.
Note that the inequality
$\sin\hat{\theta}+\sin\theta_n \le \sin\hat{\theta}(\pm\cos\theta_n)+\sin\theta_n(\pm\cos\hat{\theta})$ holds regardless of the signs of 
$\cos\hat{\theta}$ and $\cos\theta_n$, because $\sin\hat{\theta},\sin\theta_n \le 0$ and $\cos\hat{\theta},\cos\theta_n\in[-1,1]$. This concludes the proof. 
}

\vspace{-0.25cm} 
\subsection{Proof of Lemma~\ref{thm:rateReal}}
In light of Lemma~\ref{lem:negEq}, we can consider only positive $\omega_p$ without loss of generality. 
Furthermore, as a consequence of results on finite Blaschke products~\cite{Garcia_etal:BlaschkeProduct}, 
given an $n^{\rm th}$-order real-rational all-pass function $H$ with unit norm and only real poles,
$H$ can be factored into $H(s) = c H_1(s)...H_n(s)$, where 
$H_i(s)=\frac{a_i-s}{a_i+s}$ with $a_i>0$, $i=1,\cdots,n$, and $c \in \{-1, 1\}$. 

First consider $c = 1$. Let $\omega_p>0$ and $\theta_i:=\angle H_i(j\omega_p)$. One can readily verify that 
$$
H_i(j\omega_p) = \frac{a_i-j\omega_p}{a_i+j\omega_p}
=\frac{(a_i^2-\omega_p^2)-j(2a_i\omega_p)}{\omega_p^2+a_i^2},
$$
and therefore $\sin\theta_i =\frac{-2a_i\omega_p}{\omega_p^2+a_i^2}<0$. This implies $\theta_i\in(-\pi,0)$. 
Moreover, as $\|H_i\|_{H_{\infty}}=1$, we have 
$$
M_{H}'(\omega_p)=\sum_{i=1}^n M_{H_i}'(\omega_p) = \sum_{i=1}^n \frac{-2a_i}{\omega_p^2+a_i^2}
=\sum_{i=1}^n\frac{\sin\theta_i}{\omega_p}.
$$
Thus, 
\begin{align*}
\hspace{-5mm}
M_{H}'(\omega_p)
\le \frac{-|\sin(\theta_1+\theta_2+\cdots+\theta_n)|}{\omega_p}
=-\left|\frac{\sin(\theta_{H}(\omega_p))}{\omega_p}\right|,
\end{align*}
where the inequality follows from Lemma~\ref{lem:ineq_sin} and $\omega_p>0$, and the last equality follows from
$\theta_{H}(j\omega_p) = \sum_{i=1}^n \theta_i$.

Next consider $c=-1$. In this case we have
\begin{align*}
M_{H}'(\omega_p) =
M_{-H}'(\omega_p) \le -\left|\sin(\theta_{H}(\omega_p))/\omega_p\right|.
\end{align*}
{  Here we use the fact that $\sin(\theta_{-H}(\omega_p))=\sin(\theta_{H}(\omega_p))$.}
It is clear that the inequality~\eqref{eq:RateChangeReal_p1} becomes equality if the order of $H$ is one. 
It is also clear that, for the zeroth-order all-pass functions $H(s)=1$ and $H(s)=-1$, $M_H'(\omega_p) = 0$ 
and $\sin(\theta_H(\omega_p)) = 0$. So the equality also holds when the order of $H$ is zero.

\vspace{-0.25cm} 
\subsection{Proof of Lemma~\ref{thm:c<r}}
Without loss of generality, let us consider the generic rational second-order stable all-pass function 
$H(s)=\frac{s^2-bs+a}{s^2+bs+a}$ with unit gain. Since $H$ is stable, we have
$a>0$ and $b>0$. Furthermore, since $|H(j\omega)|=|-H(j\omega)|$, the theorem statement remains 
correct for the case where one or both of $H_r$ and $H_c$ have DC gain equal to $-1$. One can readily verify the
following formulas:
\begin{align*}
&H(j\omega) = \frac{((a-\omega^2)^2-b^2\omega^2)-j(2(a-\omega^2)b\omega)}{(a-\omega^2)^2+b^2\omega^2},\\
&\cos(\theta_H(\omega)) = \frac{\alpha^2-\beta^2}{\alpha^2+\beta^2}, \ \ 
\sin(\theta_H(\omega)) = \frac{-2\alpha\beta}{\alpha^2+\beta^2},
\end{align*}
where $\alpha:=a-\omega^2$, $\beta := b\omega$. Also we note that whenever $a\not = \omega^2$,
\begin{align}
\begin{split}
&M_H'(\omega) 
= \frac{\sin(\theta_H(\omega))}{\omega}+\frac{2\omega\sin(\theta_H(\omega))}{a-\omega^2}.
\end{split}
\label{eq:ChangeRate}
\end{align}

Now let $H_r(s)=\frac{s^2-b_rs+a_r}{s^2+b_rs+a_r}$, and 
$H_c(s)=\frac{s^2-b_cs+a_c}{s^2+b_cs+a_c}$. We require $a_r>0, \ b_r>0, \ a_c>0, \ b_c>0$, 
and $a_r\le b_r^2/4$, $a_c> b_c^2/4$ so that both $H_r$ and $H_c$ are stable, the former has real poles, 
and the latter complex conjugate poles. 

By Lemma~\ref{lem:negEq}, we can assume, without loss of generality, that $\omega_p > 0$.
First, consider the case where $a_r-\omega_p^2 > 0$.  
In this case, $\sin(\theta_{H_r}(\omega_p)) = \sin(\theta_{H_c}(\omega_p))$ implies that 
the two pairs $(a_r, b_r)$ and \ $(a_c, b_c)$ must satisfy 
\begin{align}
\hspace{-7mm}
\frac{2(a_r-\omega_p^2)b_r\omega_p}{(a_r-\omega_p^2)^2+b_r^2\omega_p^2} 
= \frac{2(a_c-\omega_p^2)b_c\omega_p}{(a_c-\omega_p^2)^2+b_c^2\omega_p^2} := \gamma.
\label{eq:gamma}
\end{align}
The equality leads to 
\begin{align}
\begin{split}
&(a_r -\omega_p^2 - \bar{\gamma}\omega_p b_r)(a_r -\omega_p^2 - \underline{\gamma}\omega_p b_r) = 0; \\ 
&
(a_c -\omega_p^2 - \bar{\gamma}\omega_p b_c)(a_c -\omega_p^2 - \underline{\gamma}\omega_p b_c) = 0,  
\end{split}\label{eq:LinearCons}
\end{align}
where $\bar{\gamma} = \frac{1}{\gamma}(1+\sqrt{1-\gamma^2})$ and 
$\underline{\gamma}=\frac{1}{\gamma}(1-\sqrt{1-\gamma^2})$. Note that both $\bar{\gamma}$ and $\underline{\gamma}$ are positive. 
Thus both $(a_r, b_r)$ and $(a_c, b_c)$ need to satisfy one of the following affine equations:
$a = \omega_p^2 +\bar{\gamma}\omega_p b$, and $a = \omega_p^2 +\underline{\gamma}\omega_p b$. Moreover, 
for $\cos(\theta_{H_r}(\omega_p)) = \cos(\theta_{H_c}(\omega_p))$ to hold, 
$(a_r, b_r)$ and $(a_c, b_c)$ must satisfy \emph{the same} affine equation, which then leads to 
$\theta_{H_r}(\omega_p) = \theta_{H_c}(\omega_p)$. 

Now suppose $(a_r, b_r)$ and $(a_c, b_c)$ satisfy $a = \omega_p^2 +\bar{\gamma}\omega_p b$. The inequality $a_r \le b_r^2/4$ 
implies that
\begin{align}
\begin{split}
&(b_r-2\bar{\gamma}\omega_p)^2 \ge 4(1+\bar{\gamma}^2)\omega_p^2 \\ 
  \Longleftrightarrow \ & a_r-\omega_p^2 \ge 2\bar{\gamma}(\bar{\gamma}+\sqrt{1+\bar{\gamma}^2})\omega_p^2.
\end{split}
\label{eq:ineq_ab}
\end{align}
On the other hand, the inequality $a_c > b_c^2/4$ implies that
$0<a_c-\omega_p^2 < 2\bar{\gamma}(\bar{\gamma}+\sqrt{1+\bar{\gamma}^2})\omega_p^2$. 
Putting them together yields $a_r - \omega_p^2 > a_c - \omega_p^2>0$.
In view of~\eqref{eq:ChangeRate} and {  noting} that 
$2\omega_p\sin(\theta_{H_r}(\omega_p))=2\omega_p\sin(\theta_{H_c}(\omega_p))<0$, we conclude that 
inequality~\eqref{ineq:c<r} holds 
because the second term on the right-hand side of \eqref{eq:ChangeRate} is less negative for $H_r$.  
Apparently the same arguments will hold if both $(a_r,b_r)$ and $(a_c,b_c)$ satisfy the other affine equation, 
as one simply replaces $\bar{\gamma}$ by $\underline{\gamma}$.

If $\omega_p$ is such that $a_r-\omega_p^2<0$, similar arguments will lead to inequality \eqref{ineq:c<r}. In this case,
$\gamma$ in \eqref{eq:gamma} is negative, the affine equations derived from \eqref{eq:LinearCons} are replaced by 
$a=\omega_p^2-\bar{\gamma}\omega_p b$ and $a=\omega_p^2-\underline{\gamma}\omega_p b$, 
where the forms of $\bar{\gamma}$ and $\underline{\gamma}$ are defined as those below \eqref{eq:LinearCons} 
but with $\gamma$ replaced by $|\gamma|$. The inequalities for $a_r$ and $a_c$ (stated in \eqref{eq:ineq_ab}) become
\begin{align}
\begin{split}
&a_r - \omega_p^2 \le -2\bar{\gamma}(\bar{\gamma}+\sqrt{1+\bar{\gamma}^2})\omega_p^2 <0, \\
&0 > a_c - \omega_p^2 > -2\bar{\gamma}(\bar{\gamma}+\sqrt{1+\bar{\gamma}^2})\omega_p^2.
\end{split}
\end{align}
Now with $2\omega_p\sin(\theta_{H_r}(\omega_p))=2\omega_p\sin(\theta_{H_c}(\omega_p))>0$, we again have \eqref{ineq:c<r}. 

Finally, if $a_r-\omega_p^2=0$, we have $a_c-\omega_p^2 = 0$ and 
$M_{H_c}'(\omega_p) = -4/b_c, \; M_{H_r}'(\omega_p) = -4/b_r$. 
Again, we conclude inequality in \eqref{ineq:c<r} because $-4/b_c < -2/\sqrt{a_c} = -2/\sqrt{a_r} \le -4/b_r$. 

\vspace{-0.25cm} 
\subsection{Proof of Lemma~\ref{thm:bound}}
As $\theta_f(\omega)$ and $\theta_f'(\omega)$ are both scale invariant, we can assume without loss of generality 
that $\|f\|_{H_{\infty}}=1$. 
We proceed to show that
\begin{align} 
\label{eq: bound} 
\begin{split}
&|\omega_p| \theta_f^\prime(\omega_p) \leq -\left|\theta_f(\omega_p) \right| \\
&= -\frac{2|\omega_p|}{\pi} \left| \int_0^\infty
    \frac{A_f(\omega) - A_f(\omega_p)}{\omega^2 - \omega_p^2} \, d\omega \right| .
\end{split}
\end{align}
Since $\|f\|_{H_{\infty}}=1$, we have $A_f(\omega) \leq 0$ for all $\omega \in \IR$ and $A_f(\omega_p) = 0$. 
Suppose $\omega_p\not = 0$. By Lemma~\ref{lem:CR_int_relations}, we have
\[
|\omega_p| \theta_f^\prime(\omega_p) 
= \frac{2|\omega_p|}{\pi}\int_0^\infty A_f(\omega) \frac{\omega^2 + \omega_p^2}{(\omega^2 - \omega_p^2)^2} \, d\omega.
\]
Since $\frac{\omega^2 + \omega_p^2}{(\omega^2 - \omega_p^2)^2} \geq \left|\frac{1}{\omega^2 - \omega_p^2} \right|$ for
$\omega \in [0, \omega_p) \cup (\omega_p, \infty)$, it follows 
\begin{align*}
&\hspace{-5mm}
|\omega_p| \theta_f^\prime(\omega_p) \le 
\frac{2|\omega_p|}{\pi} \int_0^\infty (A_f(\omega)-A_f(\omega_p)) \left|\frac{1}{\omega^2 - \omega_p^2} \right| \ d\omega\\
& =  -\frac{2|\omega_p|}{\pi} \int_0^\infty \left|\frac{A_f(\omega)-A_f(\omega_p)}{\omega^2 - \omega_p^2}\right| d\omega \\
& \le -\frac{2|\omega_p|}{\pi} \left| \int_0^\infty \frac{A_f(\omega)-A_f(\omega_p)}{\omega^2 - \omega_p^2} d\omega \right|.
\end{align*}
Thus~\eqref{eq: bound} holds.
Moreover, when $\omega_p=0$, by Lemma~\ref{lem:CR_int_relations} we have 
$\displaystyle
\theta_f^\prime(\omega_p) 
= \frac{2}{\pi}\int_0^\infty A_f(\omega){  /\omega^2}  \, d\omega.
$
Since $A_f(\omega)\le 0$, we conclude that $\theta_f^\prime(\omega_p)\le 0$.

\begin{IEEEbiography}[{\includegraphics[width=1in,height=1.25in,clip,keepaspectratio]{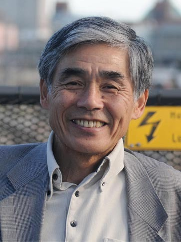}}]
{Shinji Hara}
received the B.S., M.S., and Ph.D. in engineering from Tokyo Institute of Technology 
(TITech), Japan, in 1974, 1976, and 1981, respectively. In 1984, he joined TITech as an Associate 
Professor and served as a Full Professor for ten years. From 2002 to 2017 he was a Full Professor 
in the Dept. of Information Physics and Computing at the University of Tokyo (U Tokyo). 
He is Professor Emeritus of TITech and U Tokyo. His current research interests include 
glocal control and system biology.  
Dr. Hara has received many awards in control including 
the George S. Axelby Outstanding Paper Award from the IEEE Control System Society (CSS) in 2006. 
He was the President of SICE, Japan in 2009, a Vice President of the IEEE CSS in 2009 to 2010, and 
an IFAC Council member from 2011 to 2017. He is a Fellow of IFAC, IEEE, and SICE. 
\end{IEEEbiography}

\vspace{-10mm}

\begin{IEEEbiography}[{\includegraphics[width=1in,height=1.25in,clip,keepaspectratio]{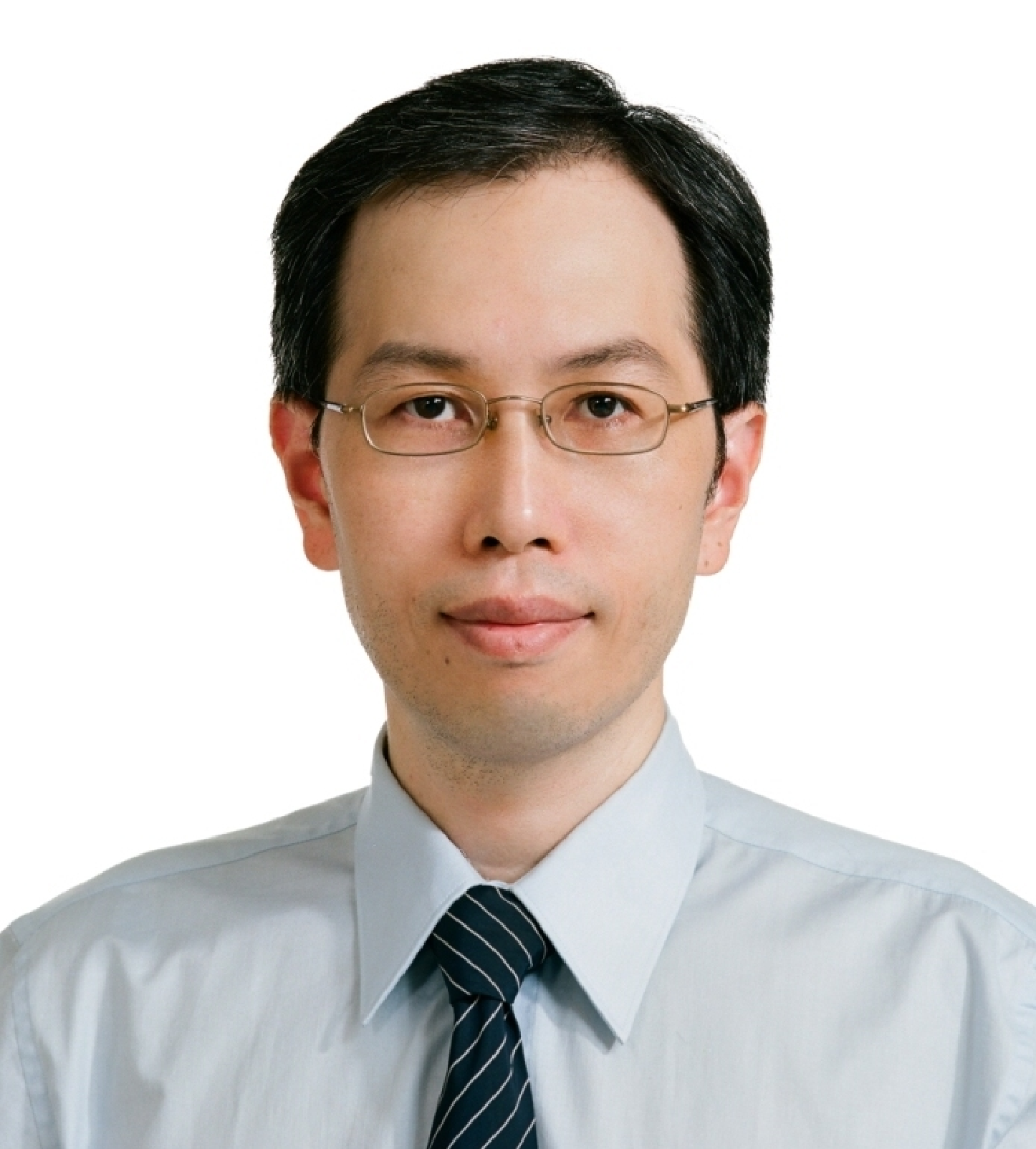}}]
{Chung-Yao Kao} 
received the Sc.D. degree in mechanical engineering from the Massachusetts Institute of Technology, Cambridge, USA, in 2002.
From 2002 to 2004, he held postdoctoral research positions at the Department of Automatic Control, Lund University, Sweden, the 
Mittag-Leffler Institute, Sweden, and the Division of Optimization and 
systems Theory, Royal Institute of Technology, Sweden. From July 2004 to January 2009, he was Senior Lecturer in 
the Department of Electrical and Electronic Engineering, University of Melbourne, Australia. In February 2009, he 
joined the Department of Electrical Engineering, National Sun Yat-Sen University, Kaohsiung, Taiwan, where he is 
currently a Professor. His research interests include robust control, systems theory, and optimization. 
\end{IEEEbiography}

\vspace{-10mm}

\begin{IEEEbiography}[{\includegraphics[width=1in,height=1.25in,clip,keepaspectratio]{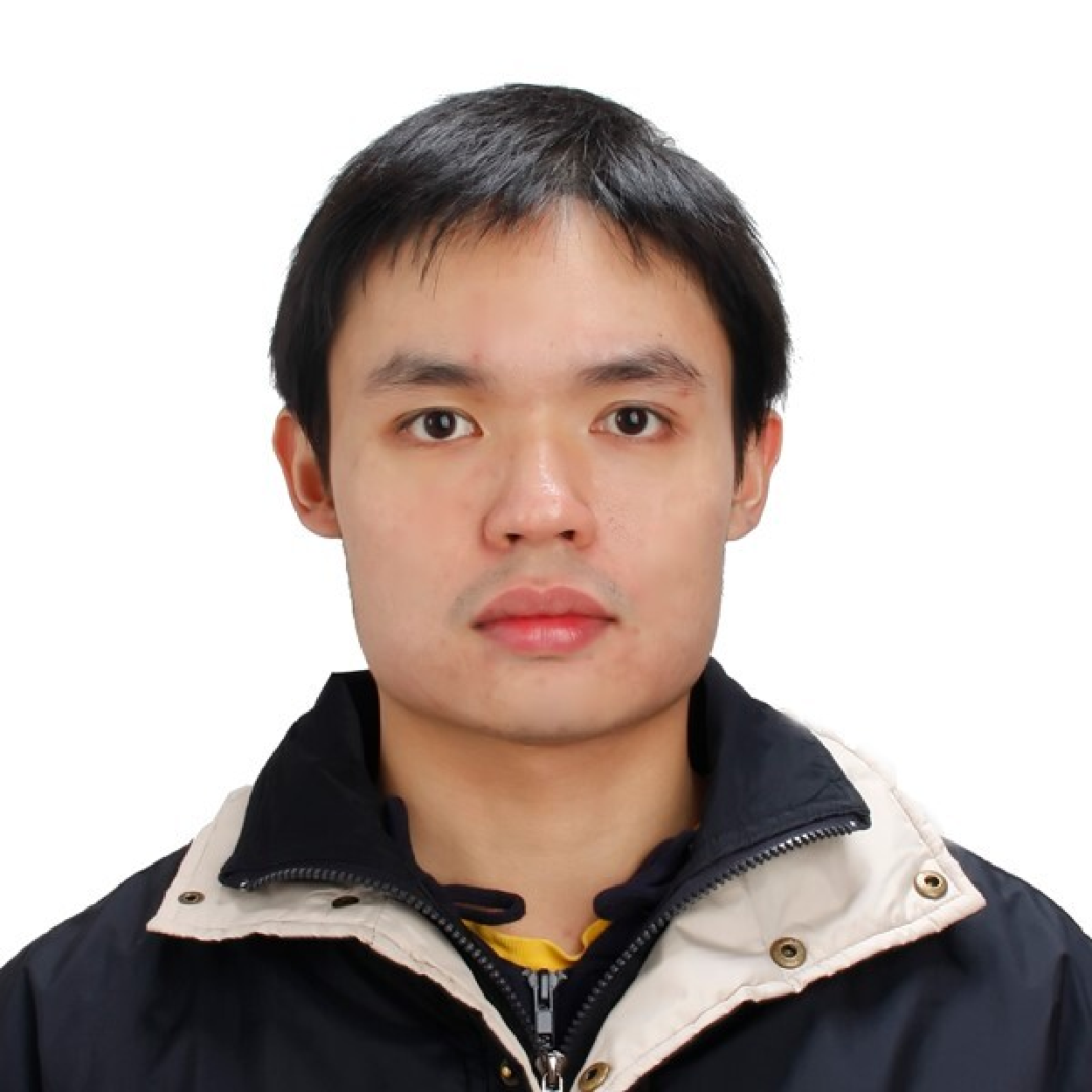}}]
{Sei Zhen Khong}
received the Bachelor of Electrical Engineering degree (with first class honours) and the Ph.D. degree from The 
University of Melbourne, Australia,
in 2008 and 2012, respectively. He has held research positions at the Department of Electrical and Electronic Engineering, 
The University of
Melbourne, Australia, the Department of Automatic Control, Lund University, Sweden, the Institute for Mathematics and 
its Applications, The University
of Minnesota, Twin Cities, USA, and the Department of Electrical and Electronic Engineering, The University of Hong Kong, 
China. His research interests
include network control, robust control, systems theory, and extremum seeking control.
\end{IEEEbiography}

\vspace{-10mm}

\begin{IEEEbiography}[{\includegraphics[width=1in,height=1.25in,clip,keepaspectratio]{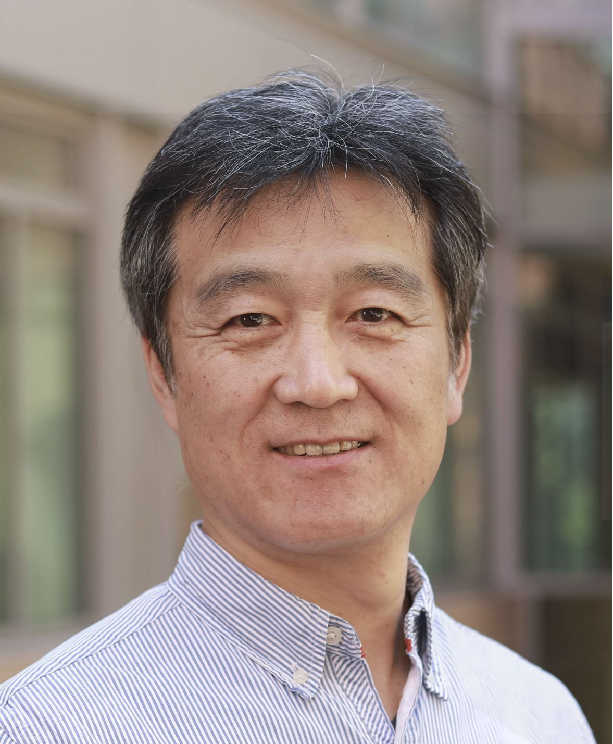}}]
{Tetsuya Iwasaki} (M'90-SM'01-F'09) received his B.S.\ and M.S.\
degrees in Electrical Engineering from the Tokyo
Institute of Technology in 1987 and 1990, 
and his Ph.D.\ degree in Aeronautics and Astronautics from Purdue
University in 1993. He held faculty positions at Tokyo Tech and
University of Virginia before joining the UCLA.
His research interests include dynamics and control
of neuromechanics, global pattern formation via local interactions, and
robust/optimal control theories. 
He has received several awards from NSF, SICE, IEEE, and ASME.
He has served as Senior/Associate Editor of several control journals.
\end{IEEEbiography}

\vspace{-10mm}

\begin{IEEEbiography}[{\includegraphics[width=1in,height=1.25in,clip,keepaspectratio]{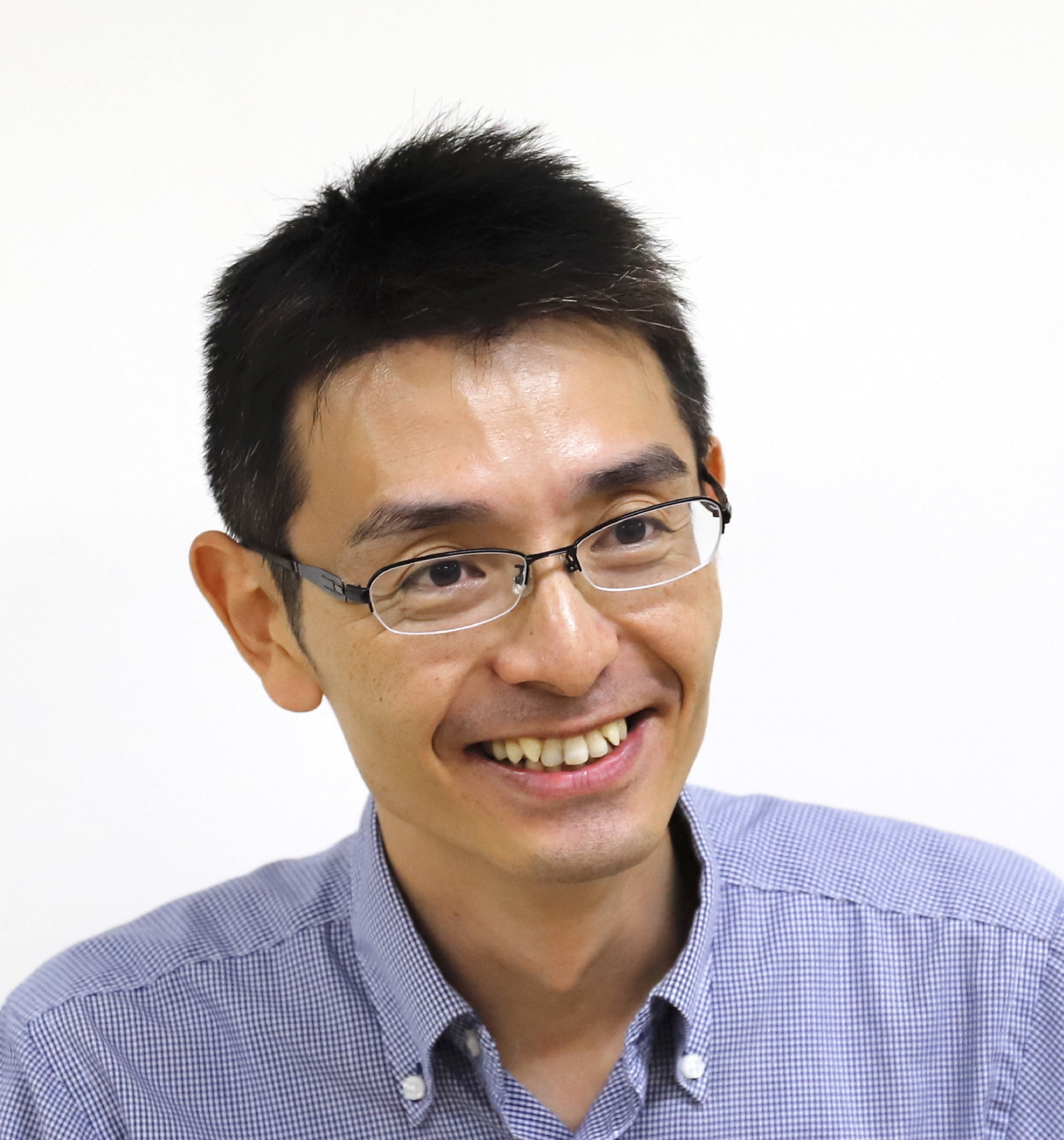}}]
{Yutaka Hori}
received the B.S degree in engineering, and the M.S. and Ph.D. degrees in information science and technology from the University of Tokyo in 2008, 2010 and 2013, respectively. He held a postdoctoral appointment at California Institute of Technology from 2013 to 2016. In 2016, he joined Keio University, where he is currently an associate professor. His research interests lie in feedback control theory and its applications to synthetic biomolecular systems. He is a member of IEEE, SICE, and ISCIE. 
\end{IEEEbiography}

\end{document}